\documentstyle[aps,epsfig,prb,multicol]{revtex}

\relax   
\begin{document}
\preprint{} \draft

\title{Instrumentation for Millimeter-wave Magnetoelectrodynamic Investigations of
Low-Dimensional Conductors and Superconductors}

\author{Monty Mola and Stephen Hill\cite{email1}}
\address{Department of Physics, Montana State University, Bozeman, MT 59717}

\author{Philippe Goy\cite{email2}}
\address{AB Millim$\grave{e}$tre, 52 Rue Lhomond, 75005 Paris, France}

\author{Michel Gross\cite{email3}}
\address{Laboratoire KASTLER-BROSSEL, C.N.R.S. UMR 8552, and Universi$t\acute{e}$ Pierre-et-Marie
Curie, D$\acute{e}$partement de Physique de l'Ecole normale
sup$\acute{e}$rieure, 24 Rue Lhomond, 75231 Paris, Cedex 05,
France}

\date{\today}
\maketitle

\begin{abstract}

We describe instrumentation for conducting high sensitivity
millimeter-wave cavity perturbation measurements over a broad
frequency range (40 $-$ 200 GHz) and in the presence of strong
magnetic fields (up to 33 tesla). A Millimeter-wave Vector Network
Analyzer (MVNA) acts as a continuously tunable microwave source
and phase sensitive detector (8 $-$ 350 GHz), enabling
simultaneous measurements of the complex cavity parameters
(resonance frequency and $Q-$value) at a rapid repetition rate
($\sim10$ kHz). We discuss the principal of operation of the MVNA
and the construction of a probe for coupling the MVNA to various
cylindrical resonator configurations which can easily be inserted
into a high field magnet cryostat. We also present several
experimental results which demonstrate the potential of the
instrument for studies of low-dimensional conducting systems.

\end{abstract}

\begin{multicols}{2}[]
\centerline{\bf I. Introduction}
\smallskip

Anisotropy, or low-dimensionality, plays a fundamental role in
many areas of contemporary condensed matter physics, both pure and
applied. For example, much of the technology which has
revolutionized the microelectronics industry in recent years has
resulted from the development of low-dimensional semiconductor
devices. More recently, research into bulk layered materials such
as the transition metal oxides,\cite{Oxides} organic
conductors,\cite{Yamaji} semiconductor superlattices,\cite{SLs}
magnetic nanostructures,\cite{multilayers} $etc$., has resulted in
the discovery of a range of new physical phenomena, $e.g$. high
temperature superconductivity,\cite{HTS} colossal
magnetoresistance,\cite{CMR} and a novel form of the quantum Hall
effect.\cite{Hillqhe} Many of these discoveries challenge our
basic understanding of condensed matter physics while, at the same
time, they hold the key to future technologies.

The use of microwave techniques to probe the electrodynamic
properties of metals is not new. However, due to recent interest
in a range of novel conducting systems in which the characteristic
energy scales ($e.g$. electronic bandwidths, or energy gaps)
coincide with the millimeter and sub-millimeter-wave spectral
ranges, this field is undergoing a renaissance. Frequency
dependent transport measurements in the presence of (strong)
magnetic fields $-$ the topic we call magnetoelectrodynamics, in
analogy of magnetooptics $-$ are expected to yield significant new
insights into the physics of low-dimensional conductors, just as
conventional magnetooptics and magnetotransport measurements were
essential decades ago in leading to our present understanding of
simple metals.


To date, few experiments on highly conducting materials have
combined high magnetic fields and millimeter-wave
spectroscopy.\cite{hillspie} The reasons for this can be
attributed to a range of factors. First, there are few
commercially available sources and detectors working in this
frequency range and those that can be used tend be highly priced,
cumbersome to use, and possess either low power or an unstable
output. Second, spectroscopy in this frequency range is complex,
particularly in the case of metals: high sample reflectivity
requires extreme sensitivity in order to detect small changes in
conductivity; and the radiation wavelengths are typically larger
than available samples, thus rendering conventional reflection
spectroscopy useless. The inclusion of a magnetic field
complicates these techniques considerably, since good optical
coupling between the spectrometer (source and detector) and the
sample under investigation is essential, and this is obviously
difficult to achieve through the small bore (comparable to
$\lambda$) of a large magnet.

In this paper we describe instrumentation which has been developed
at Montana State University (MSU) for carrying out sensitive
spectroscopy of low-dimensional conductors in the frequency range
between 40 and 200 GHz, and in magnetic fields up to 33 tesla
which are available at the National High Magnetic Field Laboratory
(NHMFL) in Tallahassee, FL. In order to a) maximize sensitivity,
and b) achieve the required control over the electromagnetic field
distribution in the vicinity of the sample, we utilize an enclosed
cavity perturbation technique (section
II-B).\cite{hillspie,cavpert} The wide frequency coverage and
outstanding dynamic range (signal-to-noise) are achieved using a
Millimeter-wave Vector Network Analyzer
(MVNA)\cite{goy,hillthesis,marcthesis} as a source and detector
(see section II-A). The resulting instrumentation offers unique
experimental possibilities in the area of metals physics in high
magnetic fields.

The paper is organized as follows: section II contains a detailed
technical description of the system; in section III, we discuss
the performance of the instrument; and we conclude with a summary
of the paper in section IV.

\bigskip

\centerline{\bf II. TECHNICAL DESCRIPTION}
\smallskip
\leftline{\bf A. The MVNA}
\smallskip

As a source and detector, we employ a Millimeter-wave Vector
Network Analyzer (MVNA)\cite{goy,hillthesis,marcthesis} to monitor
the phase and amplitude of millimeter-wave radiation transmitted
through a resonant cavity containing the sample under
investigation. The MVNA 8-350, which employs purely solid-state
electronics, allows measurements over an extended frequency range
(8 - 350 GHz)\cite{to1thz} through harmonic multiplication of a
sweepable centimeter source, $S_1$, which provides a nominally
flat output power in the range $F_1$ = 8 - 18 GHz. The mm-wave
signal used in our experiments is extracted from a Shottky diode
(harmonic generator $-$ HG) which has been optimized to produce
the desired harmonic $N$ of the sweepable cm source, $i.e$.
$F_{mm}$ = $N$ $\times$ $F_1$. Detection is then achieved by
mixing the mm-wave signal ($F_{mm}$) with the signal from a second
cm source, $S_2$, at a second Schottky diode (harmonic mixer $-$
HM). The beat frequency, $F_{beat}$, which preserves the phase and
amplitude information of the mm-wave signal relative to the local
oscillator $S_2$, is then sent to a heterodyne vector receiver
(VR).

The low noise floor of the MVNA is achieved by defining the
frequency difference between the two cm sources, $S_1$ and $S_2$,
using a main oscillator. If one then uses the same harmonic rank
on the source side (HG), and on the detection side (HM), the phase
noise associated with the cm sources cancels in the beat signal
($F_{beat}$) which is sent to the VR. Thus, the phase reference of
the VR can be taken directly from the main
oscillator.\cite{patent}

With the noise characteristics of the analyzer optimized, its
dynamic range is limited only by the harmonic conversion
efficiency of the Schottky diodes. At MSU, three pairs of Schottky
diodes are used, operating in the V ($\sim45 - 70$ GHz, $N = 3$
and 4), W ($70 - 110$ GHz, $N = 5$ and 6) and D-bands ($110 -
>200$ GHz, $N = 7$ to 12). The V-band diodes are nominally flat
broadband, while the W and D band diodes are mechanically tunable
and require optimization each time the source frequency ($F_1$) is
changed. Operating in this mode, it is possible to perform
bench-top tests up to ~350 GHz. Table I lists the optimum dynamic
ranges achieved at MSU (MVNA $8-350-1-2$)\cite{mvnatypes} in each
frequency band, and for each harmonic, up to 200 GHz ($N$ = 12).
Similar analyzers are in use at several high magnetic field
laboratories around the world;\cite{mvnas} in particular, the
national magnet labs in the US and The Netherlands have MVNA 8-350
analyzers (with ESA options)\cite{to1thz} which are available to
external users. The MVNA at the NHMFL, which has been used for
some of the studies discussed in this paper, additionally operates
in the Q ($30 - 50$ GHz, $N = 3$), K ($16 - 32$ GHz, $N = 2$) and
X ($8 - 18$ GHz, $N = 1$) bands.

Although the cm sources ($S_1$ and $S_2$) are phase (frequency)
locked to each other, their absolute frequencies must be
stabilized also. The frequency precision and stability provided by
the MVNA 8-350 is not adequate for narrow-band cavity perturbation
measurements when the bandwidth of the cavity is less than about
100 MHz. For this reason, it is common to phase-lock one of the
sources ($i.e.$ both) to a quartz standard. At MSU and at the
NHMFL, EIP 575 source-locking frequency counters\cite{EIP}  are
used, which provide both the stability and precision necessary for
the measurements described in this paper. One other mode of
operation involves phase locking the source ($S_1$) directly to
the high$-Q$ cavity resonator used for the experiment. The counter
is also useful in this case for recording changes in frequency
resulting from any changes in dispersion within the cavity.
Comparisons between the two frequency locking techniques are
discussed in section III-B.

The dynamic ranges listed in Table I represent ideal values for
the MVNA 8-350 assuming there are no insertion losses between HG
and HM. Due to the considerable size of a typical magnet cryostat
and, in particular, the dimension separating the magnetic field
center and the top of the cryostat ($\sim 1.5$ m in our case), a
considerable insertion loss is unavoidable. Furthermore, it is
essential to keep a reasonable distance ($\sim 2$ m) between the
MVNA and the magnet, since the cm sources (YIG oscillators) are
sensitive to stray magnetic fields of more than a few Gauss. The
Schottky diodes (HG and HM), on the other hand, are not field
sensitive and may be placed closer to the experiment. Thus, the cm
and beat frequencies are propagated between the MVNA and Schottky
diodes through flexible coaxial cables, which introduce a combined
insertion loss of 4 to 5 dB. Meanwhile, the mm-wave signal is
propagated from the HG to the cavity within the magnet cryostat,
and back to the HM, using a pair of rigid waveguides. These
waveguides account for a major part of the insertion loss of the
system; minimizing these losses is the subject of section II-B.1.
A schematic of this arrangement is shown in Fig. 1.

In addition to its use as a source and detector for solid state
spectroscopy, the network analysis capability of the MVNA is
crucial during the developmental stages of a measurement system.
The frequency sweeping capability can be used to determine the
precise location of impedance mis-matches and/or microwave leaks
from the system, and for the characterization and optimization of
different resonator designs.

\bigskip

\leftline{\bf B. The sample probe}
\smallskip

As outlined in the introduction, our aim is to be able to conduct
sensitive cavity perturbation measurements, at low temperatures,
high magnetic fields and over a broad frequency range. In this
section, we describe an experimental scheme $-$ the "sample probe"
$-$ for coupling the MVNA to various enclosed cylindrical
resonator configurations which can easily be inserted into a high
field magnet cryostat. The "sample probe" refers to the passive
microwave hardware which has been developed at MSU and is
compatible with the MSU and NHMFL magnet systems. It consists of
the following components: two long waveguides for propagating
mm-waves from the HG, into the magnet cryostat (incidence
waveguide), and back to the HM (transmission waveguide); a
demountable cavity, of which we have developed two standard types;
a coupling between the cavity and the incidence and transmission
waveguides; a vacuum tube to isolate the probe from the
surrounding liquid cryogens; and electronics for controlling the
sample temperature, and for magnetic field modulation. Fig. 1
shows a schematic of the probe situated within the superconducting
magnet cryostat at MSU, and Fig. 2 shows an equivalent LRC circuit
representation of the key microwave components; we defer
explanation of these figures until following sections, where
various aspects of the probe design are discussed in detail.

From the outset, we established two key experimental objectives:
i) it should be possible to make measurements at several well
separated frequencies without the need to interfere with, or warm
up, the sample probe; and ii) the sensitivity, dynamic range and
mechanical stability of the measurement system should be optimized
as far as possible. The flexibility in choice of frequency is
important for experiments at the NHMFL, where magnet time is often
limited. There are also other advantages to this, $e.g$. by not
having to interfere with the sample, it is possible to ensure that
it sits in an identical electromagnetic field distribution for a
given series of cavity modes. The second goal is fairly self
explanatory, nevertheless, high sensitivity comes at the expense
of some flexibility. We have, therefore, opted for a rigid
construction, $i.e$. no in-situ rotation of either the cavity or
the sample within the cavity is possible. Our experience has shown
that such mechanisms result in radiation leakage from the cavity
and diminished sensitivity.

\bigskip

\leftline{\bf\it 1. Reducing losses in the waveguides}
\smallskip

The success of our recent measurements owes as much to reductions
in the insertion losses associated with our sample probe, as it
does to the optimization of the cavity designs. In this section,
we discuss the technical details of how to propagate mm-wave
radiation from the Schottky diodes into a low temperature cryostat
(pumped $^4$He) within the bore of a high field magnet system
where the resonant cavity containing the sample is situated. As
illustrated in Fig. 1, the MSU setup utilizes a separate inner
cryostat to control the temperature of the experiment. This
cryostat, which sits within the bore of the superconducting
magnet, draws liquid helium from the main reservoir into its tail,
which is thermally isolated from the main helium reservoir, thus
allowing good control over the sample temperature.

The microwave probe is based on V-band rectangular waveguides
which are terminated at their source/detector ends with UG385/U
flanges for convenient connection to the Schottky
diodes.\cite{penn} V-band waveguide cuts off below $\sim 45$ GHz,
which is sufficiently low for most of the intended applications,
and is also the lowest frequency band currently available at MSU.
Furthermore, the narrow bores of the magnets at the
NHMFL\cite{bore} essentially rule out the possibility of using
waveguides below Q-band (cut-off at 33 GHz) for transmission
measurements.

Commonly used low loss, high conductivity, waveguide materials
such as copper or silver are not ideal for our purposes.
Unacceptable heat flow down the waveguides and into the cryostat
leads to excessive liquid helium boil-off and limits the ability
to cool the cavity/sample to pumped $^4$He temperatures ($\sim
1.5$ K). For cryogenic purposes, Stainless Steel (SS) waveguide
offers an attractive alternative:\cite{penn,atwall} it has both a
low thermal conductivity,\cite{kappa} and it is possible to use
thinner walled waveguide material.\cite{wall} Unfortunately the
microwave losses in SS waveguides are severe, as illustrated in
Fig. 3, which shows a comparison between the insertion losses of
SS and Ag, as measured using the frequency sweeping capability of
the MVNA;  note that, in this range, the loss in SS exceeds 10
dB/m, whereas the loss in Ag is less than 2 dB/m.\cite{rho} The
solution, therefore, is to construct composite waveguides from
highly conducting waveguide sections isolated by SS sections.
Although this necessitates several joints in the waveguide
assembly, we have found that these do not significantly impair the
functionality of the waveguide.

Placement of the SS waveguide sections is crucial to the cryogenic
performance of the system, as illustrated in Fig. 1. The largest
temperature gradient occurs in the upper part of the magnet dewar
where the radiation baffles are situated. It is in this region
that we first use SS. We again use short SS sections at the lower
end of the probe in order to thermally isolate the cavity/sample
from the long Cu/Ag\cite{cuag} sections which run from just below
the baffles, through the liquid $^4$He reservoir, and into the
bore of the magnet. The cavity and sample are, therefore,
more-or-less isolated from the main 4.2 K helium reservoir,
enabling control over the temperature of the cavity/sample from
about 1.35 K, up to 40 or 50 K (see section II-B.4). The lower SS
waveguide sections have been gold plated\cite{cmi} for reasons
which are discussed in the following section.

One unforeseen benefit of the composite waveguide construction,
which greatly improves the mechanical stability of the probe, is
its insensitivity to the liquid helium level in the magnet
cryostat. We assume that the Cu/Ag section maintains a fairly
uniform temperature ($\sim 4.2$ K) over its entire length due to
its high thermal conductivity. Thus, the temperature gradient in
the SS section does not depend on the liquid helium level. In
contrast, the thermal gradient in a single SS section immersed in
liquid helium would depend on the level of the liquid, causing the
waveguide to expand slightly over time, thereby affecting the
microwave phase stability of the system.

The SS and Cu/Ag rectangular waveguide sections, which have
different outer dimensions,\cite{wall} are coupled together using
specially machined clamps which screw tightly around the
waveguides.\cite{cam} The joints are staggered so as to minimize
cross talk between the waveguides due to any microwave leakage
from the joints (see following section). Thus, although both
waveguides pass through each clamp, only one of the waveguides is
joined at each clamp. This construction is extremely rugged, yet
it easily permits modifications in the overall probe length, and
in the placement of the SS sections. Therefore, the same waveguide
assembly can easily be re-configured for use at other magnet
facilities such as the NHMFL.

Finally, the entire sample probe fits tightly inside a 19.05mm OD
(0.4mm wall) vacuum jacket. The integrity of the vacuum is
maintained across the waveguides at their room temperature ends
using mylar windows clamped between standard flanges (UG385/U)
which have been modified to hold rubber 'O'-rings. Again, we
offset the vacuum seals on the incident and transmission
waveguides, and wrap them in steel wool to minimize cross-talk. A
hermetically sealed connector provides 19 electrical feedthroughs
for thermometry, field modulation coils, and for powering a
heater.

\bigskip

\leftline{\bf\it 2. Minimizing undesirable instrumental effects }
\smallskip

The basic principal behind the cavity perturbation technique is
relatively simple, and involves measuring changes in the complex
cavity parameters ($i.e$. resonance frequency $f_o$ and $Q-$value)
upon insertion of a sample.\cite{hillspie,cavpert} Relating these
changes to the complex electrodynamic properties of the sample can
be a formidable task, and we refer the reader to a series of three
articles\cite{klein,donovan,dressel} dealing with this analysis
for precisely the types of materials which we set out to study in
high magnetic fields, $i.e$. highly anisotropic crystalline
conductors. However, in order to reliably apply such an analysis
in our case, it is essential that we first consider how each
element in the sample probe affects our ability to extract the
relevant information from the cavity mounted at the end of the
probe. As we shall show, this is a formidable task in itself,
which is complicated considerably by the restricted access into
the bore of a typical high field magnet cryostat, and by the need
for the probe to accommodate cavities resonating over a broad
range of frequencies.

The task of optimizing the probe design is best tackled by
considering an equivalent AC circuit (Fig. 2). Each component of
the probe may be modeled as a self contained LRC circuit which is
inductively coupled to the next. The Schottky diodes, HG and HM,
attach to the upper ends of the incidence and transmission
waveguides, which have impedances {\bf Z}$_I$ and {\bf Z}$_T$,
respectively. The coupling between HG (HM) and the incidence
(transmission) waveguide is modeled as a coupling mutual
inductance {\bf $m_{HG}$} ({\bf $m_{HM}$}). The cavity, which is
mounted at the lower end of the two waveguides and has impedance
{\bf Z}$_C$, is coupled to each of the waveguides through coupling
mutual input ({\bf $m_{in}$}) and output ({\bf $m_{out}$})
inductances. Ideally, the incidence and transmission waveguides
should be coupled solely through the cavity. However, a leak
signal between the incidence and transmission waveguides, in
parallel with the cavity, is inevitable; this is modeled as a
direct coupling mutual inductance, {\bf $m_{l}$}, between the two
waveguides.

It is changes in {\bf Z}$_C$, caused by the insertion of a sample
into the cavity, which one would like to measure in a cavity
perturbation experiment. Since the MVNA measures phase and
amplitude, it is important to see how {\bf Z}$_C$ affects both of
these parameters. Before, considering the probe as a whole, $i.e$.
the entire circuit in Fig. 2, we examine the cavity by itself.
Variations in the amplitude and phase of a wave transmitted
through a cavity, as the frequency ($f$) is swept across a
resonance, are given by

\begin{eqnarray}
A^{2}(f)= \frac{1}{1 + [2 (f - f_o)/\Gamma]^2}\label{e.1}
\end{eqnarray}
\leftline{and}
\begin{eqnarray}
\phi(f)= -\arctan \frac{2 (f - f_o)}{\Gamma}, \label{e.2}
\end{eqnarray}

\noindent where $f_o$ is the center frequency and $\Gamma$ is the
FWHM of the resonance; these expressions have been normalized so
that $A$ = 1 and $\phi$  = 0 at resonance. As $f$ is swept from 0
to $\infty$, $A$ and $f$ sweep out a circle in the complex plane,
where the real and imaginary amplitudes are given by $A_1(f)=A
\cos \phi$ and $A_2(f)=A \sin \phi$ respectively, or

\begin{eqnarray}
A_1(f)= \frac{1}{1 + [2 (f - f_o)/\Gamma]^2}\label{e.3}
\end{eqnarray}
\leftline{and}
\begin{eqnarray}
A_2(f)= \frac{2 (f - f_o)/\Gamma}{1 + [2 (f - f_o)/\Gamma]^2},
\label{e.4}
\end{eqnarray}

\noindent the more familiar expressions for a Lorentzian. Such a
circle is shown in Fig 4a, for a resonance at 60 GHz, with a $Q$
of 5000. Each point in the figure corresponds to a different
frequency, and the frequency interval between each point is 800
kHz. Points closest to the origin correspond to $f\ll$ 60 GHz and
$f\gg$ 60 GHz. The resonance frequency ($f_o$) lies along the real
axis ($\phi = 0$) where the spacing between points is greatest,
$i.e$. when the phase rotates most rapidly with frequency.
Clockwise rotation around the circle represents increasing
frequency. The corresponding changes in $A(f)$ \& $\phi(f)$ and
$A_1(f)$ \& $A_2(f)$ are shown in Figs 4b and c, respectively.

In a locked frequency experiment, changes in dispersion (Im\{{\bf
Z}$_C$\}) cause the cavity to go off resonance, $i.e$. the point
corresponding to $f_o$ in Fig. 4a will move away from the real
axis. Provided that this effect is weak (a perturbation), the
dominant result is a change in the phase of the signal transmitted
through the cavity, and no appreciable change in amplitude.
Conversely, changes in dissipation (Re\{{\bf Z}$_C$\}) within the
cavity will result in a reduction in the amplitude of the signal
transmitted through the cavity and, hence, a reduction in the
diameter of the circle in Fig. 4a. Dissipation alone does not move
$f_o$ away from the real axis and, therefore, does not affect the
phase of the wave. However, dispersion can affect both the
amplitude and the phase of the wave if $f_o$ moves appreciably off
the real axis. For this reason, it is often desirable to conduct a
locked phase experiment, which completely decouples these two
effects. With a phase lock, the cavity stays on resonance (on the
real axis in Fig. 4a), and dispersion affects the resonance
frequency only, which we can measure with a frequency counter.
Meanwhile, changes in dissipation again affect the amplitude of
the transmitted signal only.

Unfortunately, each of the additional circuit elements required to
link the MVNA to the cavity, and/or improper coupling between
these components, has the potential to seriously distort the
simple relationships between dissipation, dispersion, amplitude
and phase discussed above. Ideally, the sample probe should be
passive, low loss, insensitive to temperature and magnetic field
and, with the exception of the cavity, should have a flat
broad-band frequency response. In practice, this is never actually
possible to achieve. Nevertheless, by conducting a thorough
characterization and optimization of each element in the microwave
circuit (Fig. 2), it is possible to minimize these instrumental
effects to negligible levels. The MVNA performs a pivotal role in
this hardware developmental process. The following paragraphs
discuss various undesirable instrumental characteristics, their
potential effect on a measurement, and the steps we have taken to
eliminate these sources of error.

\smallskip

$A$ $leak$ $wave$ bypassing the cavity directly through to the
transmission waveguide has two adverse effects. First, it
diminishes the useful dynamic range $-$ ideally 100\% of the
signal reaching the detector should pass through the cavity.
Second, if the leak amplitude is comparable to the amplitude of
the signal passing through the cavity, the resonance may become
severely distorted, making it extremely difficult to distinguish
between dissipative and dispersive effects within the cavity. As
illustrated in Fig. 5, a leak wave adds a complex vector to the
signal transmitted through the cavity. By minimizing the leak, one
can control its amplitude. However, it is not possible to control
the phase of the leak wave. Consequently, big leaks lead to an
arbitrary vector translation of the circle in Fig. 4a; this is a
pure translation, $i.e$. each point on the circle is translated by
the same vector. Hence, the line joining the resonance frequency,
$f_o$, and the $f = 0$ \& $\infty$ points, remains parallel to the
real axis. As a result, the transmitted amplitude on resonance is
not necessarily the maximum amplitude; indeed, it can take on any
value from zero to one plus the leak amplitude. This is
illustrated in Fig. 5 for an arbitrary translation of the circle
in Fig. 4a, together with the corresponding variations in phase
and amplitude, plotted versus frequency. It is apparent from this
figure that the phase of the leak signal is entirely responsible
for the way in which dissipation and dispersion affect the phase
and amplitude of the signal transmitted through the cavity. Thus,
an appreciable leak signal is intolerable, and we have taken every
step to reduce the leak in our probe to at least 20 dB below the
typical signal transmitted through the cavity on resonance, as
discussed above.

\smallskip

$Standing$ $waves$ in the waveguides are unavoidable and, without
proper attention, can cause considerable problems, especially when
operating in the phase-locked mode in which the incident mm-wave
frequency is locked to the cavity resonance frequency. Changes in
this frequency will result in changes in the phase and amplitude
of the mm-wave signal incident upon the cavity. Thus, it becomes
impossible to distinguish between the intrinsic cavity response
and spurious effects due to the standing waves. Furthermore, the
phase is no longer truly locked to the cavity resonance under
these circumstances, but rather to the coupled response of the
entire circuit in Fig. 2. Standing waves should not be ignored
altogether in the frequency locked mode either, particularly when
the cavity is well coupled to the waveguides. Under these
circumstances, changes in {\bf Z}$_C$\ influence the impedance
matching between the cavity and the waveguides ($i.e$. {\bf
$m_{in}$} and {\bf $m_{out}$}) and, therefore, affect the standing
waves.

There is little to be gained from trying to eliminate the standing
waves completely. This would require precise impedance matching of
each component in Fig. 2, which is only possible to achieve over a
narrow frequency range and would, therefore, defeat the purpose of
the probe, which is intended to work over a fairly broad frequency
range. Instead, we concentrate on minimizing the influence of the
standing waves on a measurement at any given frequency. This is
achieved by reducing the frequency bandwidth of the measurement to
well below the periodicity of the standing wave pattern, $i.e$. so
that the response of the waveguides is essentially flat over the
relevant frequency interval. In this way, a range of cavities or
cavity modes may be utilized, covering an extremely broad
frequency range in comparison to the standing wave periodicity.
Meanwhile, each cavity mode samples only a minute portion of the
waveguide spectrum, over which its response is essentially flat.

The fastest standing wave period is governed by the longest
dimension of the sample probe, which is ~3m in our case (HG to HM,
via the cavity). This gives rise to a standing wave periodicity of
about 100 MHz which, in turn, requires cavity filling factors of
less than $10^{-4}$, so that the frequency shift in any given
measurement never exceeds about $10^{-4}$ of the measurement
frequency, $i.e$. $\Delta f_o < 10$ MHz for $f < 100$ GHz. Cavity
filling factors of between $10^{-5}$ and $10^{-4}$ are typical for
the types of samples we study, so standing wave problems do not
force this restriction upon us. Nevertheless, to compensate for
the small filling factors, it is essential to have cavity
$Q-$factors on the order of $10^{4}$. In order to attain such high
$Q-$values, the cavity must necessarily be coupled weakly to the
waveguides, which further reduces problems associated with
standing waves.


\smallskip

$Phase$ $instability$, and/or a strong frequency dependence of the
phase reaching the VR, can lead to a variety of problems. Although
the internal sources within the MVNA are phase locked, additional
phase jitter can arise in the mm-wave signal due to mechanical
and/or thermal instabilities in the sample probe. Mechanical
vibrations do not pose any problems, either at MSU or at the
NHMFL. However, thermal stability is essential for achieving the
best results. Due to extremely high cavity $Q-$values, minor
temperature fluctuations can lead to significant phase
instabilities, particularly when studying samples with a strongly
temperature dependent electrodynamic response. For this reason,
active temperature control is necessary (see section II-B.4). Long
term thermal stability of the entire probe, including the 1.5m
waveguide sections, is also desirable and is best achieved using
the composite waveguides described in the previous section.

A strong frequency dependence of the phase reaching the VR occurs
when the distances between the sources $S_1$ and $S_2$ and the HM
are very different. The signal originating at $S_1$ has to travel
the extra 2 $\times$ 1.5 m into, and out of, the magnet cryostat.
For this reason, it can be beneficial to compensate for the extra
distance between $S_1$ and the HM by adding 3 m of coaxial
cable\cite{phirot} between $S_2$ and the HM, especially when
performing phase-locked or frequency swept measurements, even
though this introduces an extra 2 $-$ 3 dB insertion loss.

\smallskip

$Spurious$ $magnetic$ $resonances$ caused by paramagnetic
contamination of either the cavity, or the waveguides close to the
magnetic field center, will give rise to both sharp and broad
instrumental features (Electron Paramagnetic Resonances - EPR) in
the transmission versus magnetic field response of the system.
These spurious resonances can be hard to distinguish from the
genuine response of the sample within the cavity, and should be
eliminated so that the magnetic field response of the unloaded
probe ($i.e$. with no sample) is as flat as possible. The SS
waveguide sections cause the most severe problems, which we
suspect is due to the presence of small traces of Fe$^{3+}$ (rust)
at the surface of the metal. For this reason, we have gold plated
the SS sections which couple directly to the cavity. This not only
reduces the insertion loss due to these sections, but also
completely eliminates a broad instrumental resonance, leaving an
extremely flat response ($\Delta A < 0.05$ dB), as shown in Fig.
6. A number of other measures have been taken to avoid
contaminating the lower end of the probe ($i.e$. close to the
field center) with paramagnetic impurities. These include clamping
all components together rather than using adhesives or solders,
and avoiding the use of cutting oils when machining the
resonators. We have also gold plated some of the cavities to
prevent tarnishing. Nevertheless, electrolytic OFHC Cu cavities
work extremely well without plating (see Fig. 6).

\bigskip

\leftline{\bf\it 3. The cavities}
\smallskip

To date, our efforts have focussed on the use of enclosed
cylindrical resonators for cavity perturbation measurements. There
are several compelling reasons to do so. To begin with, the simple
design concept we have developed, in which the cavities are
assembled from relatively few easily machinable components (see
Figs. 7 and 8), enables us to fabricate a range of cavities and,
therefore, switch frequency range/coverage with relative
ease.\cite{ease} More importantly, relative to rectangular
cavities, it is straightforward to achieve extremely high
$Q-$values ($>10^4$) for the TE$01n$ ($n=1, 2, 3...$) modes of
cylindrical resonators. The reason for this has to do with the
fact that no AC currents flow between the end and side walls of a
cylindrical cavity excited in its TE$01n$ mode. Consequently,
joints at these ideal locations (see Figs 7 and 8) do not diminish
the cavity $Q-$factor. These modes also possess electromagnetic
field geometries which are highly desirable for the measurements
described in this paper.

\bigskip

\leftline{\bf\it 3.1. Axial cavities}
\smallskip

For the most part, we use axial cavities of the type shown in Fig.
7, $i.e$. ones in which the cavity axis is coincident with the
axis of the superconducting solenoid. This is by far the most
versatile design, though it does have some limitations, especially
when it is necessary to tilt the sample with respect to the
applied DC magnetic field. The cavity is constructed from OFHC
electrolytic copper in three pieces - a blank end plate, a
coupling plate and a cylindrical barrel. This assembly bolts onto
the under side of a copper housing which clamps around the
incidence and transmission waveguides. These bolts also provide
the pressure for reproducibly clamping the cavity assembly tightly
together; no adhesives or solders are used. The cavity may be
disconnected from the waveguides in a matter of seconds for easy
sample insertion/removal. The housing which clamps around the
waveguides is also easily interchanged with a separate housing for
a transverse cylindrical cavity (see below).

Coupling between the waveguides and the cavity is achieved by
means of small circular apertures in the thin coupling plates.
Since these plates terminate the waveguides and the cavity, there
are no transverse microwave electric ({\bf\~ E}-) fields at the
locations of the apertures. Thus, it is the microwave magnetic
({\bf\~ H}-) fields in the waveguides and the cavity which should
be matched. For the TE$01n$ modes, the {\bf\~ H}-fields flow
radially at the cavity ends and, therefore, the {\bf\~ H}-fields
in the waveguides should do so also. For this reason, the incoming
waveguides are oriented with their shortest edges closest together
(see Figs 1, 7 and 8).\cite{poole}

We can control the degree of coupling between the waveguides and
the cavity ($m_{in}$ and $m_{out}$) by means of the dimensions of
the coupling apertures (diameter and thickness). There is an
important trade off here between i) strong coupling (large
apertures), which ensures good power throughput from the source to
the detector and, hence, a large dynamic range, and ii) weak
coupling (small apertures), which limits radiation losses from the
cavity, resulting in higher cavity $Q-$values and increased
sensitivity, at the expense of some dynamic range. We have found
empirically that the optimum coupling apertures should be small
for our setup, with diameter $\sim \lambda /4$. It is also
necessary for the coupling plate to be very thin ($\sim \lambda
/20$), since the signal is obviously attenuated as it passes
through the apertures, which are way below cut-off. The relatively
weak coupling to the cavity makes it all the more important to
reduce any/all other losses in the sample probe and, hence,
preserve a reasonable dynamic range, $i.e$. without taking steps
to reduce losses in the waveguides (see section II-B.1), we would
be forced to increase the coupling, resulting in reduced
sensitivity.

The resonance frequency of a particular cavity is determined by
the length and the inner diameter of the cylindrical barrel (see
Fig. 7). Because of the simplicity of machining this section, it
is no great task to construct a large number of cavities,
providing many TE$01n$ modes covering a wide range of frequencies.
However, because of the wavelength dependence of the coupling
between the waveguides and the cavity, it is also necessary to
construct several different coupling plates $-$ roughly one for
each frequency band, $i.e$. V-, W-, D-, $etc$.. Because the inner
diameters of the cavities generally get smaller at higher
frequencies, the positioning of the apertures is also critical. In
all cases, the input and output coupling apertures are located
diametrically opposite each other, and at the same radial distance
from the axis of the cavity. In order to minimize the number of
these coupling plates, we have limited the cavity diameters to
four standard sizes. For V-band (cavity dia. $= 9.52$ mm), the
coupling apertures are located at the optimum positions, both with
respect to the cavity and the waveguides, $i.e$. in the middle of
the waveguides and half way between the cavity axis and its
perimeter. For the higher frequency bands, a compromise between
these positions is made.

A clear gap of 0.51mm is maintained between the incidence and
transmission waveguides, which terminate flush with the under side
of the cavity housing. The coupling plate is machined with a
0.51mm wide, 0.75mm high, ridge running perpendicular to the line
joining the coupling apertures, and intersecting the mid point
between them (see Fig. 7). This ridge locates inside a matching
groove on the under side of the cavity housing and, therefore, in
between the incidence and transmission waveguides, as shown in
Fig. 7. We have found that it is this joint between the cavity and
the waveguides which is most susceptible to microwave leaks, and
that the ridge in the coupling plate dramatically reduces the leak
amplitude. The leak may be reduced further still by applying a
small amount of Indium into the groove before assembling the
cavity.


For a perfectly cylindrical cavity, the TE$01n$ modes are
degenerate with the TM$11n$ modes. It is a trivial task to lift
this degeneracy without diminishing the $Q-$values of the TE$01n$
modes. We achieve this by drilling a small indent in the center of
the blank end plate where essentially no currents flow for the
TE$01n$ modes; this hole may also be used to hold a quartz pillar
for mounting a sample on axis above the cavity end plate (see Fig.
9). Consequently, the TM modes are shifted to lower frequencies by
up to 200 MHz, they are also weaker and have lower $Q-$values than
the TE modes.

\bigskip

\leftline{\bf\it 3.2 Transverse cavities}
\smallskip

Recently, we have developed a transverse cavity optimized to work
in V-band (shown in Fig. 8). Although the construction of this
cavity is a little more complex than the axial cavities, it offers
a major advantage for experiments in which it is necessary to tilt
the sample relative to the applied DC magnetic field. The major
difference between this cavity and the axial one is that the
mm-waves are coupled through apertures drilled directly through
the side wall of the cylindrical barrel, $i.e$. a separate
coupling plate is not employed. This requires removing material
from one side of the cylindrical barrel so that the side wall of
the cavity is sufficiently thin in the vicinity of the coupling
apertures. This flat surface also facilitates attachment to the
underside of a specially designed cavity housing, which again
clamps around the incidence and transmission waveguides. The
cavity end plates simply bolt onto either end of the cavity
barrel; one of these end plates has machined slots at the bolt
circle radius so that it may be rotated about the cavity axis.

As with the axial cavity, we again implement a ridge/groove
arrangement between the waveguides and the cavity, in order to
minimize any microwave leakage at this position. The locations of
the coupling apertures in the side wall of the cavity have been
optimized for exciting the TE$012$ mode, $i.e$. at $^1/_4 h$ and
$^3/_4 h$ from the ends of the cavity, where $h$ is the cavity
height. Nevertheless, these positions provide good coupling to the
TE$011$ mode and, to a lesser extent, the TE$013$ mode.

\bigskip

\leftline{\bf\it 3.3 Sample positioning, and sample rotation}
\smallskip

Using one or other of the two cavity types, it is possible to
subject a sample to virtually any combination of AC  {\bf\~ E}-
and  {\bf\~ H}-field polarizations, relative to the applied DC
magnetic field $({\bf B_o})$ orientation, $e.g$. at an {\bf\~ H}
node with {\bf\~ E}//${\bf B_o}$, or at an {\bf\~ E} node with
{\bf\~ H}$\bot {\bf B_o}$, $etc$.. However, it turns out that for
studies of highly anisotropic conductors, two convenient locations
are sufficient for most experiments utilizing TE$01n$
modes.\cite{hillspie,cavpert,donovan} In the "end plate"
configuration, the sample is placed on the blank cavity end plate,
exactly half way between the cavity axis and its perimeter, as
shown in Fig. 9a. In the "quartz pillar" configuration, the sample
is mounted atop a thin quartz pillar (dia. = 0.71mm) on the axis
of the cavity, as shown in Fig. 9b. In both of these
configurations, the sample sits in an {\bf\~ H}-field antinode for
the TE011 mode, the polarization of which is radial for the end
plate configuration and axial for the quartz pillar configuration
(see Figs 9a and b).

For the end plate configuration, all TE$01n$ modes have radial
{\bf\~ H}-field antinodes at the sample location. This is not the
case for the quartz pillar configuration; for example, if the
sample is mounted precisely mid-way between the cavity end plates,
the even $n$ modes have both {\bf\~ E} and {\bf\~ H}-field nodes
at this location. Careful forethought as to the positioning of the
sample can rectify this problem to a certain extent. However,
whenever positioning the sample away from the mid-point ({\bf\~
H}$_{max}$-point) of the cavity, sensitivity is compromised.
Indeed, even the end plate position is appreciably less sensitive
than the cavity mid-point.\cite{donovan}

To understand how it is that the end plate and quartz pillar
locations within a TE$01n$ cavity can be sufficient for studying
anisotropic systems, we consider the electrodynamics of a
quasi-two-dimensional (Q2D) conductor. We assume a quasi-static
approximation and consider the Faraday {\bf\~ E}$_F$-field
resulting from the time varying {\bf\~ H}-field, $i.e$. an
oscillatory {\bf\~ E}$_F$-field which curls around the
polarization of the {\bf\~ H}-field. In an isotropic conductor,
this Faraday field would induce circulating currents in a plane
perpendicular to the {\bf\~ H}-field. Indeed, this is
approximately what happens if the {\bf\~ H}-field is polarized
perpendicular to the highly conducting planes of a Q2D conductor,
as illustrated in Fig. 9c. However, because of the high in-plane
conductivity, these induced currents are damped within the
interior of the sample, $i.e$. currents only circulate within the
skin layer at the edge of the sample. Consequently, the in-plane
complex surface impedance, ${\bf Z}_S = {\bf R}_S + i{\bf X}_S$,
governs the electrodynamic response of the sample in this
situation, where ${\bf R}_S$ and ${\bf X}_S$ are the surface
resistance and reactance respectively.\cite{klein,donovan,dressel}
If, instead, the {\bf\~ H}-field is polarized parallel to the
highly conducting layers, {\bf\~ E}$_F$ will induce both in-plane
and inter-layer currents. As before, the in-plane currents will
only flow at the edges of the sample, whereas the inter-layer
currents flow throughout the bulk of the sample, due to the poor
conductivity in this direction. Consequently, the complex
inter-layer conductivity dominates the electrodynamic response of
the sample under these conditions, as depicted in Fig.
9d.\cite{klein,donovan,dressel}

Of major interest for studies of low-dimensional conductors is the
possibility to rotate the applied DC magnetic field, ${\bf B_o}$,
with respect to the sample. Since the cavities are rigidly
connected to the sample probe, the orientation of ${\bf B_o}$
cannot be adjusted relative to the cavity, $i.e$. ${\bf B_o}$ is
fixed parallel to the cavity axis for the axial cavity, and
perpendicular to the axis for the transverse cavity. Thus, the
axial cavity essentially limits investigation to only two
orientations of ${\bf B_o}$ relative to the sample, namely
parallel or perpendicular to the Q2D layers, which we denote ${\bf
B_{//}}$ and ${\bf B_{\bot}}$ respectively. One could tilt the
sample away from either of these geometries within the cavity.
However, this would result in a misalignment of the {\bf\~
H}-field polarization relative to the sample. This is the main
reason for developing the transverse cavity. Because of the
cylindrical symmetry of the TE$01n$ modes, one can rotate the
sample about the cavity axis by means of rotating the cavity end
plate to which it is attached, without affecting the polarization
of {\bf\~ H} relative to the sample; this works for both the end
plate and quartz pillar configurations, as illustrated in Fig. 9e.

\bigskip

\leftline{\bf\it 4. Temperature and magnetic field control}
\smallskip

Being constructed from a block of high conductivity copper, the
cavity makes an excellent heat reservoir for controlling the
temperature of the sample, which is kept in good thermal contact
with the cavity at all times, either by directly attaching it to
the cavity end plate, or by mounting it on a quartz pillar;
Silicone grease is used to hold the sample in place. A coil of
high resistance wire is used as a heater. This coil is wound
around a pillar which screws into the cavity housing and is,
therefore, easily interchanged between the various cavity
geometries (not shown in Figs 7 and 8). Low pressure helium gas is
admitted into the 19.05mm vacuum jacket for exchanging heat
between the sample probe and the surrounding pumped liquid helium
cryostat. Thus, the cavity and the sample can accurately and
controllably be maintained at any temperature in the range from
1.35 K up to about 50 K. The temperature is stabilized according
to a calibrated Cernox resistance thermometer embedded in the
walls of the copper cavity. Cernox thermometers have negligible
magnetoresistance above 4.2 K, and we can correct for a weak
magnetoresistance below 4.2 K.

Magnetic fields at MSU are generated using a standard commercial
superconducting solenoid and power supply. This magnet routinely
operates to 8 T, and will go to 9 T at pumped liquid helium
temperatures. At the NHMFL, the strongest magnetic fields are
produced in axial Bitter-type water cooled resistive magnets
powered by 20 MW supplies. Presently, the highest available
continuous field is 33 T.\cite{hybrid} Details of the NHMFL
magnets are published elsewhere.\cite{brooks}


\bigskip

\centerline{\bf III. PERFORMANCE }
\smallskip
\leftline{\bf A. Tests}
\smallskip

Fig. 10 shows a frequency sweep across the TE011 mode of an axial
cavity (cavity A, height = 6.7mm, dia. = 9.52mm, see Table II);
the cavity is loaded with a sample of
$\kappa$-(BEDT-TTF)$_2$Cu(SCN)$_2$ (approx. dimensions
$0.5\times0.50\times0.2$ mm$^3$)\cite{hillkappa} in the end plate
configuration, and the temperature is 4.2 K. In the upper panel,
we plot linear amplitude versus phase. The data points (squares)
form a perfect circle passing through the origin, indicating
negligible leak vector. The frequency interval between each point
is approximately 250 kHz and the solid line is a fit to the data.
In the lower panel, we plot both phase and linear amplitude
(normalized) versus frequency. This resonance is perfectly
symmetric due to the fact that the leak amplitude is 34.5 dB below
the amplitude on resonance. The loaded $Q-$value of the cavity is
19,000 and, thus, the resonance width (2.34 MHz) is considerably
less than the standing wave period, which is on the order of 100
MHz. The absolute value of the phase returned by the VR is
arbitrary,\cite{arbitrary} which is why the phase on resonance is
64$^o$ rather than $0^o$ (see Fig. 4a). In any subsequent
experiment, we would null the phase on resonance, and interpret
changes in the complex parameters of the signal returned to the VR
according to the procedure described in section II-B.2. It should
be noted that this data is about as good as one could expect were
the cavity mounted on the bench top and the HG and HM connected
directly to the cavity, $i.e$. the influence of the intervening
waveguides has been completely eliminated.

Next, we consider the influence of the nearby TM111 mode on
measurements made at the TE011 resonance frequency. Fig. 11 shows
two such resonances obtained at liquid helium temperature (cavity
A): the main part of the figure plots linear amplitude versus
frequency; the inset shows the circles in the complex plane
obtained for each of the resonances. The TM111 mode has been
shifted 230 MHz below the TE011 mode, which corresponds to almost
100 times the width of the TE011 mode (${\bf \Gamma}_{TE011}$ =
2.42 MHz) and about 40 times the width of the TM111 mode (${\bf
\Gamma}_{TM111}$ = 5.73 MHz). The resonance amplitude of the TM111
mode is about 60\% of the TE011 resonance amplitude. However, more
importantly, the power of the TM111 signal (Obtained by
extrapolation of a Lorentzian fit) is 44.5 dB below the power of
the TE011 signal when the TE011 mode is at resonance. Thus, for
all intents and purposes, we can rule out any interference between
these modes. Even if there were a slight mixing, both modes have
{\bf\~ H}-fields perpendicular to the applied DC field (${\bf
B_o}$) for the end-plate configuration, and the TM111 mode has an
{\bf\~ H}-field node at the center of the cavity where the sample
is usually placed in the quartz-pillar configuration; this has
been verified experimentally using electron paramagnetic resonance
standards.\cite{epr}

$Q-$values as high as 24,900 have been obtained at liquid helium
temperatures for the loaded axial cavities excited in TE011 modes.
In general, higher $n$ (higher $f_o$) TE$01n$ modes have reduced
$Q-$values. In addition, the shorter wavelengths associated with
the higher frequencies slightly increases the leak amplitude
relative to the signal transmitted through the cavity. These
facts, together with the diminished dynamic range of the
spectrometer (see Table I) at higher frequencies, make it harder
to observe TE$01n$ ($n > 1$) resonances of comparable quality to
the TE011 modes. Nevertheless, the data in Fig. 10 far exceeds the
criteria discussed in section II-B for making successful cavity
perturbation measurements. Consequently, we have been able to make
reliable measurements at frequencies up to 130 GHz.

Above about 130 GHz, we have less confidence in the mode
assignment of the resonances. However, by following the frequency
dependence of data containing distinct features which also depend
strongly on the polarization of the AC-fields within the cavity
(see following section), we have been able to characterize and use
axial cavity modes all the way up to 180 GHz. Fig. 12 shows
several higher $n$ TE$01n$ axial cavity modes, together with
selective higher frequency resonances. Table II lists the
frequencies, $Q-$values, leak amplitudes and dynamic ranges
associated with these modes, as well as some parameters for the
transverse cavity (all at 4.2 K). These figures clearly
demonstrate the potential of the system for cavity perturbation
measurements. It should also be noted from Fig. 12 that many of
the resonances were obtained using a single resonator (cavity A),
which was one of the main objectives for this system from the very
outset.\cite{ease} In the following section, we show real data
obtained over the frequency range covered by this single cavity.

Finally, Fig. 13 illustrates the importance of using active
temperature stabilization: the upper panel shows magnetic
resonance data taken at the base temperature of the cryostat (~1.4
K); the upper panel shows the same data obtained using active
temperature stabilization at 1.8 K. A clear drift in the unlocked
temperature data is observed throughout the course the up and down
sweeps of the magnetic field.

\bigskip

\leftline{\bf B. Experimental examples}
\smallskip

By eliminating the influence of all components of the sample probe
(aside from the cavity) on the signal returned to the VR, it is
possible to record changes in the complex cavity parameters in
real time, $i.e$. the vector (either amplitude and phase, or
amplitude and frequency) recorded at the VR is directly related to
the impedance of the cavity, ${\bf Z_C}$. Here, once again, we see
the power of the MVNA. Using a scalar detection scheme, it would
be necessary to modulate the frequency in order to extract the
complex cavity response.\cite{donovan} This would inevitably
result in a much longer time for recording each data point. The
MVNA essentially returns phase and amplitude information at the
detection frequency of the VR, which is approximately 10
kHz.\cite{goy,vr} This aspect of the instrument described in this
paper makes it highly suited to measurements in high magnetic
fields, which can be expensive to run for long periods.

A distinct advantage of conducting fixed frequency optical
measurements, as a function of magnetic field, is that the
spectral features which are under investigation may be expected to
change with field, $i.e$. the magnetic field is the variable which
is used to tune the electronic excitation spectrum of the material
under investigation. If, for example, this induces a change from
an insulating state, to a metallic state, then the optical
response at frequencies comparable to the gap will reflect this
change. A beautiful example of this can be seen from data obtained
at the NHMFL, (Fig. 14) which shows the complex electrodynamic
response of the organic superconductor (TMTSF)$_2$ClO$_4$, at low
temperatures, as the magnetic field is swept to 30
tesla.\cite{tmtpap} A rich behavior is observed; each of the more
pronounced features may be attributed to modifications in the
electronic configuration of the system which are known to occur in
the field ranges from ~5 to 10 tesla
(Field-Induced-Spin-Density-Waves - FISDW), and from ~20 to 30
tesla (a phase line within the final FISDW phase).\cite{mckernan}
Magneto-quantum-oscillations are also observed at high fields.

The data displayed in Fig. 14 were obtained using a phase lock
while recording changes in amplitude ($A$) and the resonance
frequency ($f_o$) of an axial cavity excited at 47.2 GHz. The
sample was oriented with its $\bf\em{c}$* axis parallel to the
applied DC magnetic field, ${\bf B_o}$, and currents were excited
in the $\bf\em{ab}$ plane of the sample. Each field sweep took
less than 10 minutes. The quality of this data is, at worst,
comparable to an equivalent DC measurement. From the combined
information obtained from changes in dissipation ($\Delta A$) and
dispersion ($\Delta f_o$) within the cavity, we can obtain
valuable information about the electrodynamics of the bound and
free carrier systems in this fascinating
material.\cite{tmtpap,mckernan}

The above example is quite different from a zero-field measurement
where data are usually taken at many frequencies in order to
investigate a particular spectral feature. Such experiments are
generally plagued by poor dynamic range (signal-to-noise),
especially in the millimeter and sub-millimeter spectral ranges.
This is because identical coupling between the sample and the
spectrometer cannot be guaranteed at each frequency, resulting in
a large scatter of the data. This is unfortunate, since the cavity
perturbation technique is inherently sensitive and, through the
use of a suitable spectrometer ($e.g$. the MVNA), provides plenty
of dynamic range at any given frequency. The instrument described
in this paper is, instead, optimized at each frequency in order to
detect minute changes in the optical conductivity of a sample as a
function of magnetic field. These changes may subsequently be
normalized to the zero-field conductivity using an appropriate
instrument.

Whether to choose a phase lock or a frequency lock depends on the
experiment. A frequency lock generally provides better stability
because of the higher $Q-$value of the quartz frequency reference,
as compared to the cavity. Fig. 15 shows an example of separate
measurements using both techniques. The sample under investigation
is the purple bronze $\eta
-$Mo$_4$O$_{11}$.\cite{cavpert,hillmoly2} Fig. 15a shows
variations in amplitude for both up and down sweeps of the
magnetic field. There is a slight hysteresis, which is well
documented for this material. However, the transmitted amplitude
is considerably lower at high magnetic fields for the frequency
locked measurement. This can be attributed to the cavity going
well off resonance, as evidenced by a phase change of $25^o$ in
the first 7 or 8 tesla (Fig 15b). Inspection of Fig. 4 clearly
illustrates that a $25^o$ phase shift will lead to such a
reduction in amplitude, even if there is no change in the
dissipation within the cavity. Indeed, the ratio of the amplitudes
obtained by each technique scales nicely with the cosine of the
phase shift, as shown in the inset to Fig. 15b. Consequently, the
frequency lock is inappropriate in this case, because of
considerable mixing of the dissipative and dispersive responses of
the sample in the phase and amplitude returned to the VR. Fig. 15c
shows the frequency shift observed using the phase lock.

Fig. 16 shows Periodic Orbit Resonances (PORs $-$ related to
cyclotron resonance)\cite{hillsolo} observed through the
inter-layer conductivity of the quasi-two-dimensional organic
conductor $\alpha
-$(BEDT-TTF)$_2$TlHg(NCS)$_4$.\cite{hillpphmf,hillprep} The
frequencies, which are indicated above each trace in the figure,
were obtained using a single axial cavity, and range all the way
from 44 GHz to 182 GHz. The sample was mounted in the end-plate
configuration. Not all of the frequencies correspond to TE$01n$
modes. However, from the shapes of the resonances
($\sim$Lorentzian peaks), we can be certain that the sample sits
in the same electromagnetic environment for all of the modes,
$i.e$. with the polarization of the oscillatory {\bf\~ H}-field
parallel to the conducting layers, resulting in the excitation of
inter-layer currents (see Fig. 9d). This can be confirmed by
studying the same sample in the TE011 quartz pillar configuration,
as shown in Fig. 17, which is the conventional geometry for
observing cyclotron resonance. Because of a high in-plane
conductivity, it is the in-plane surface resistance of the sample
that governs the dissipation in the cavity (see Fig.
9c).\cite{klein,dcnqi} This is the reason for the rather
unconventional lineshape, $i.e$. an inflection rather than a
symmetric dip or peak.

The data in Figs 16 and 17 are exceptional in their quality when
compared to earlier attempts to measure cyclotron resonance in
organic conductors.\cite{prevcr}  Furthermore, the resonance line
shapes agree precisely with theory, thereby providing absolute
confidence in the ability of the technique to discriminate between
in-plane and inter-layer transport phenomena in
quasi-two-dimensional conductors. This has traditionally been
problematic using conventional DC resistivity probes, because of
uncertainties in the current paths within the samples. This opens
up a huge range of possibilities for tackling issues concerning
the role of electronic dimensionality in the physical properties
of low-dimensional systems.

\bigskip
\centerline{\bf IV. SUMMARY}
\smallskip

We have described an instrument for conducting millimeter-wave
cavity perturbation measurements over a continuously tunable
frequency range (40 $-$ 200 GHz). The system is compatible with
magnets both at Montana State University (up to 9 tesla) and at
the National High Magnetic Field Laboratory (up to 33 tesla) in
Tallahassee, FL. The utilization of a Millimeter-wave Vector
Network Analyzer enables simultaneous measurements of the complex
cavity parameters (resonance frequency and $Q-$value) at a rapid
repetition rate ($\sim10$ kHz). Several experimental examples are
presented which demonstrate the potential of this system for
studying the magnetoelectrodynamics of low-dimensional conducting
systems.

\bigskip
\leftline{\bf ACKNOWLEDGEMENTS}
\smallskip

We are indebted to Norm Williams for technical assistance and to
Prof. J. S. Brooks for the use of the MVNA at the NHMFL. This work
was supported in part by the Office of Naval Research, and by NSF
cooperative agreement No. 98-71922 with the state of Montana. Work
carried out at the NHMFL was supported by a cooperative agreement
between the State of Florida and the NSF under DMR-95-27035.


\end{multicols}

\clearpage

\begin{table}
\centerline{\epsfig{figure=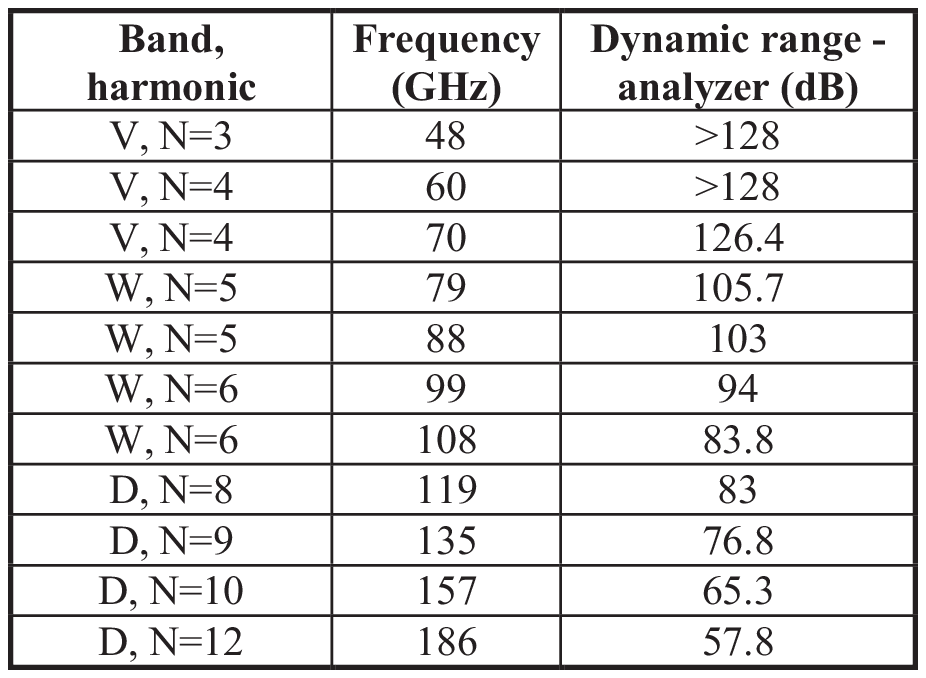,width=120mm}} \bigskip
\caption {The dynamic ranges achieved using the MVNA-8-350 at MSU
at various frequencies in each microwave band.}\label{Fig. 01}
\end{table}

\bigskip

\begin{table}
\centerline{\epsfig{figure=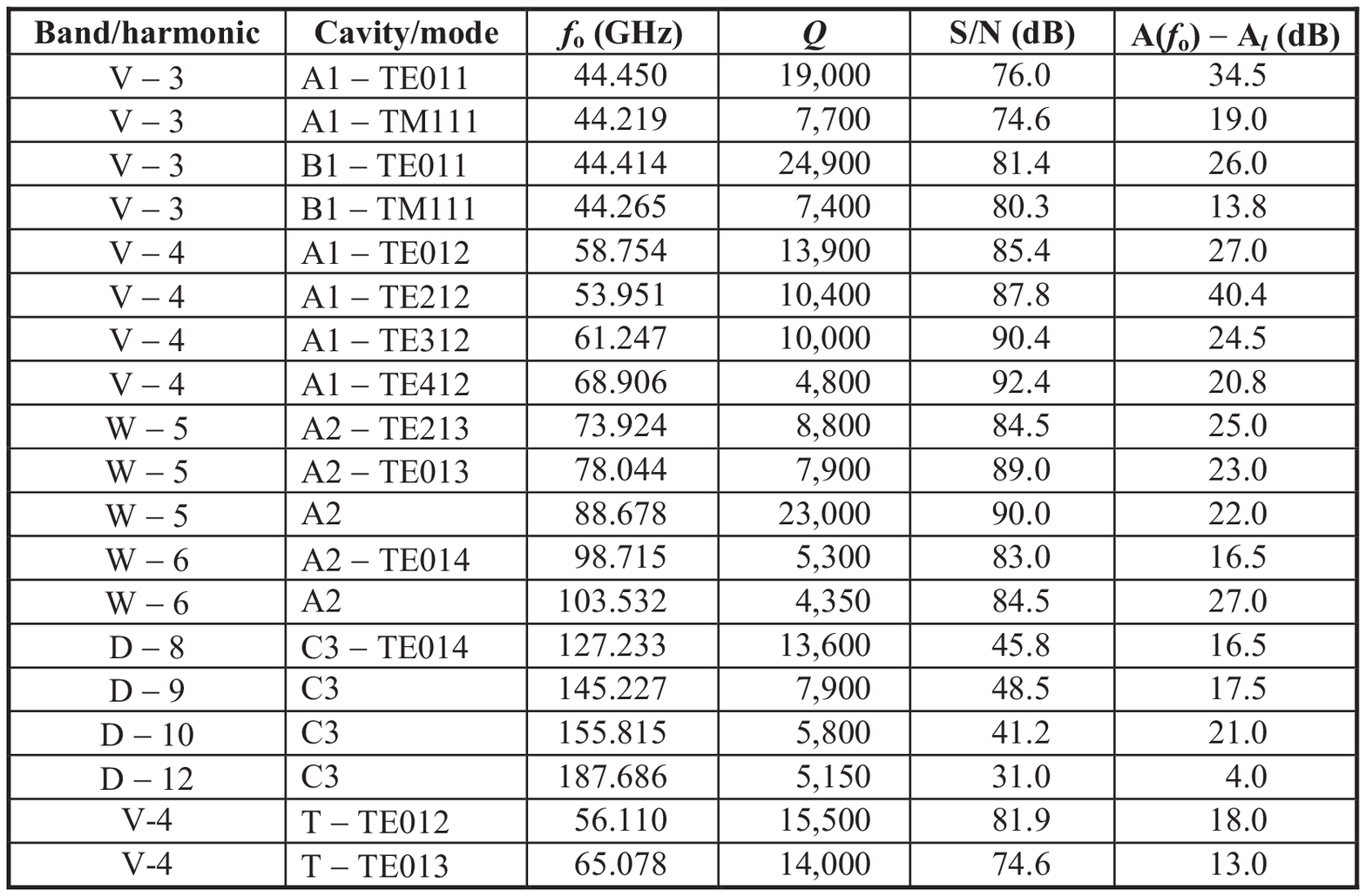,width=120mm}} \bigskip
\caption{Resonance parameters for different modes obtained in
various different cavities. The first column lists the frequency
band and the MVNA harmonic. The second column lists the cavity
(letter) and end plate used (number), as well as the mode (if
known); cavities A, B and C have dimensions (length $\times$
diameter) 6.7mm $\times$ 9.52mm, 8.73mm $\times$ 9mm, and 5.54mm
$\times$ 5.64mm respectively, cavity T is the transverse cavity
(10.16mm $\times$ 7.75mm), and end plates 1, 2 and 3 have coupling
hole diameters of 1.32mm, 0.84mm and 0.62mm, respectively. The
next three columns list the resonance frequencies ($f_o$),
$Q-$values and dynamic ranges (signal-to-noise) for each mode. The
final column lists the contrast in dB between the transmitted
amplitude on resonance [$A(f_o)$] and the leak amplitude ($A_l$).
The dynamic range is somewhat lower at 45 GHz than in the main
part of the V- and W-bands due to the fact that the V-band
wave-guide is so close to cut off at these low frequencies.}
\label{Fig. 02}
\end{table}

\clearpage
\begin{twocolumn}

\begin{figure}
\centerline{\epsfig{figure=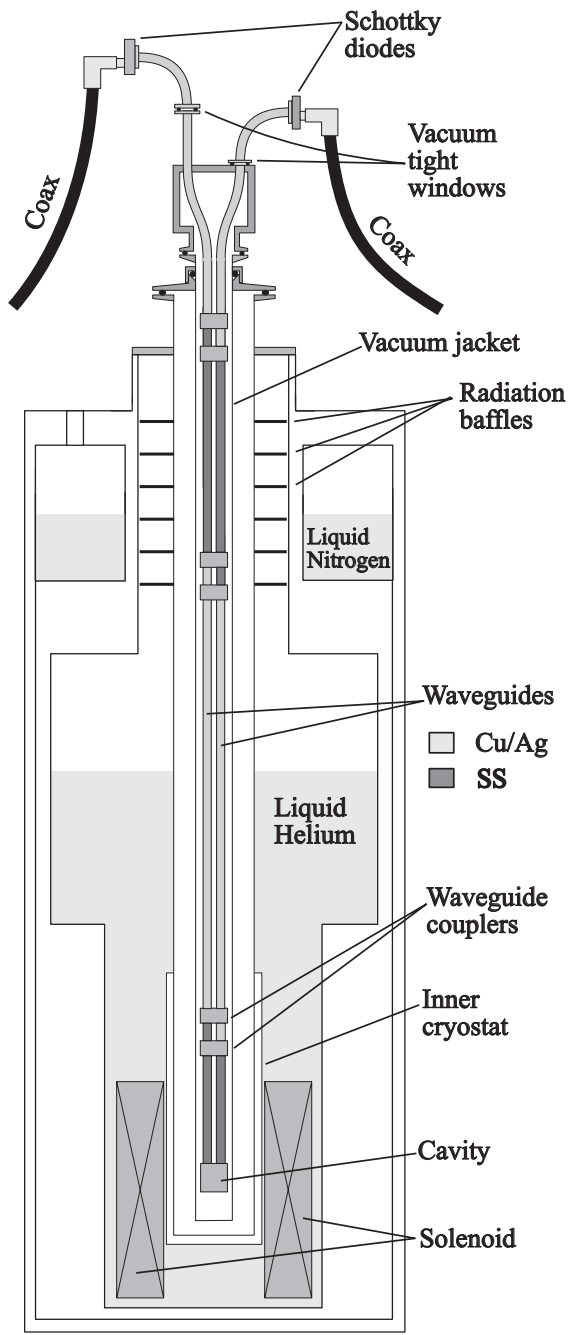,width=75mm}} \bigskip
\caption{Schematic of the sample probe and superconducting magnet
system at MSU (not to scale).} \label{Fig. 1}
\end{figure}


\begin{figure}
\centerline{\epsfig{figure=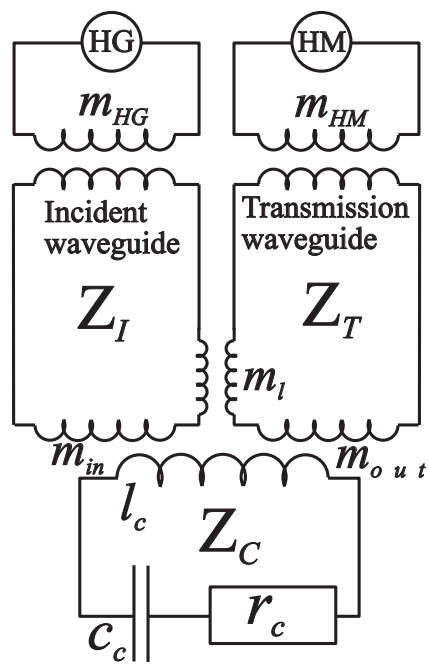,width=55mm}} \bigskip
\caption{An equivalent LRC circuit representation of the key
microwave components of the sample probe. The coupling between HG
(HM) and the incidence (transmission) waveguide is modeled as a
coupling mutual inductance $m_{HG}$ ($m_{HM}$). The cavity is
coupled to each of the waveguides through coupling mutual input
($m_{in}$) and output ($m_{out}$) inductances. A leak signal
between the incidence and transmission waveguides, in parallel
with the cavity, is modeled as a direct coupling mutual
inductance, $m_l$, between the two waveguides.} \label{Fig. 2}
\end{figure}

\bigskip


\begin{figure}
\centerline{\epsfig{figure=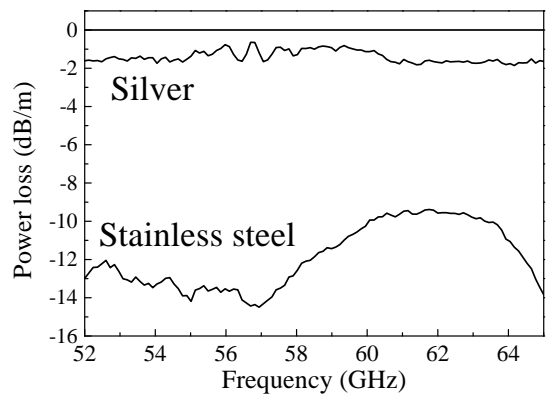,width=80mm}} \bigskip
\caption{A comparison between the insertion losses (in dB/m) for
stainless steel and Silver waveguide sections in V-band at room
temperature.} \label{Fig. 3}
\end{figure}


\begin{figure}
\centerline{\epsfig{figure=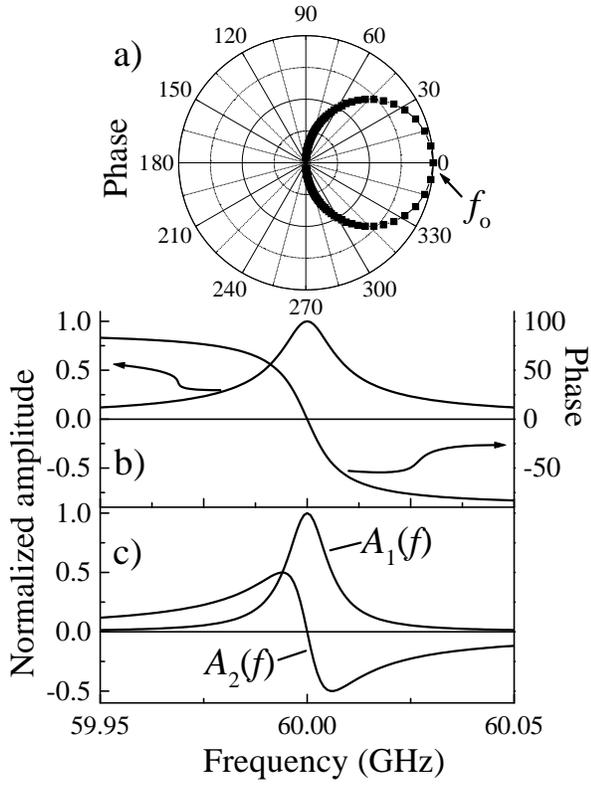,width=80mm}} \bigskip
\caption{Simulation of the complex parameters of a wave
transmitted through a cavity which is resonant at 60 GHz, and has
a $Q$ of 5000; a) a plot of linear amplitude versus phase in the
complex plane; b) a plot of linear amplitude (normalized) and
phase versus frequency; and c) a plot of the real ($A_1$) and
imaginary ($A_2$) linear amplitudes (normalized) versus
frequency.} \label{Fig. 4}
\end{figure}


\begin{figure}
\centerline{\epsfig{figure=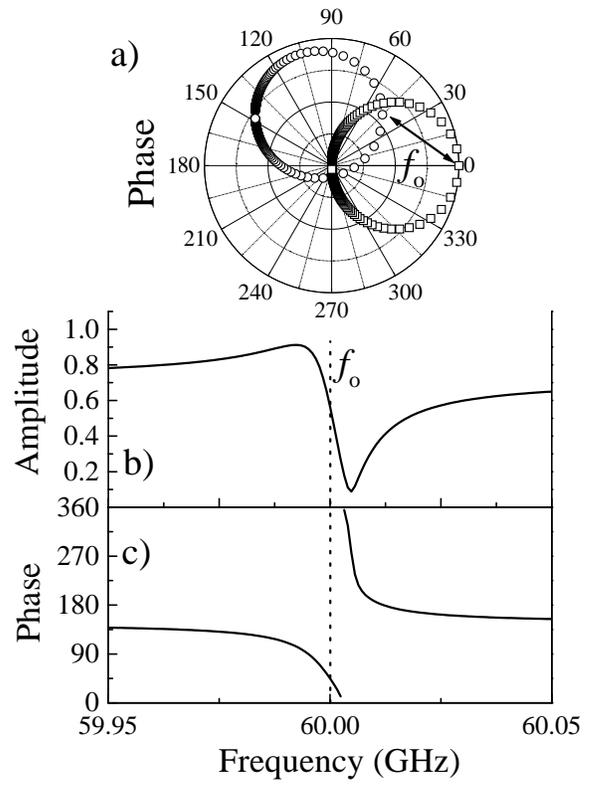,width=80mm}} \bigskip
\caption{A demonstration of the effect of a leak wave on the ideal
resonance in Fig. 4. The complex leak vector translates the
resonance in the complex plane, as shown by the arrow in a). The
result is that the resonance frequency, $f_o$, no longer lies
along the real axis ($\phi = 0$). Thus, $f_o$ no longer
corresponds to the maximum amplitude, resulting in the distorted
Lorentzian lineshape shown in b) and c).} \label{Fig. 5}
\end{figure}

\clearpage

\begin{figure}
\centerline{\epsfig{figure=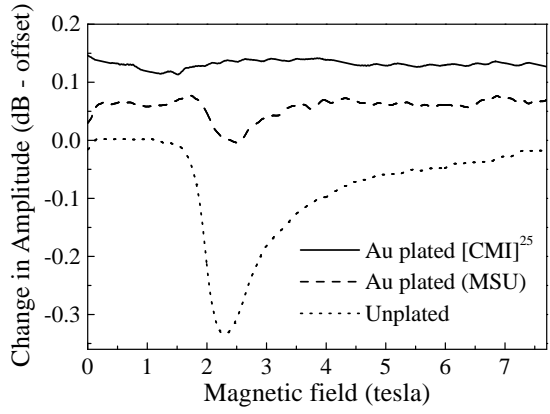,width=80mm}} \bigskip
\caption{A comparison between the background (empty Cu cavity)
response of the sample probe using Gold plated and unplated
stainless steel sections at the lower end of the probe; these data
were obtained at 76.7 GHz and 4.2 K. The broad dip is attributed
to an EPR absorption due to small quantities of Fe$^{3+}$ (rust)
at the surface of the SS. Professional gold plating (solid line)
clearly eliminates this contamination, which is evident both in
the unplated data (short dash) and the data obtained after our
efforts to plate the SS at MSU (long dash).} \label{Fig. 6}
\end{figure}


\begin{figure}
\centerline{\epsfig{figure=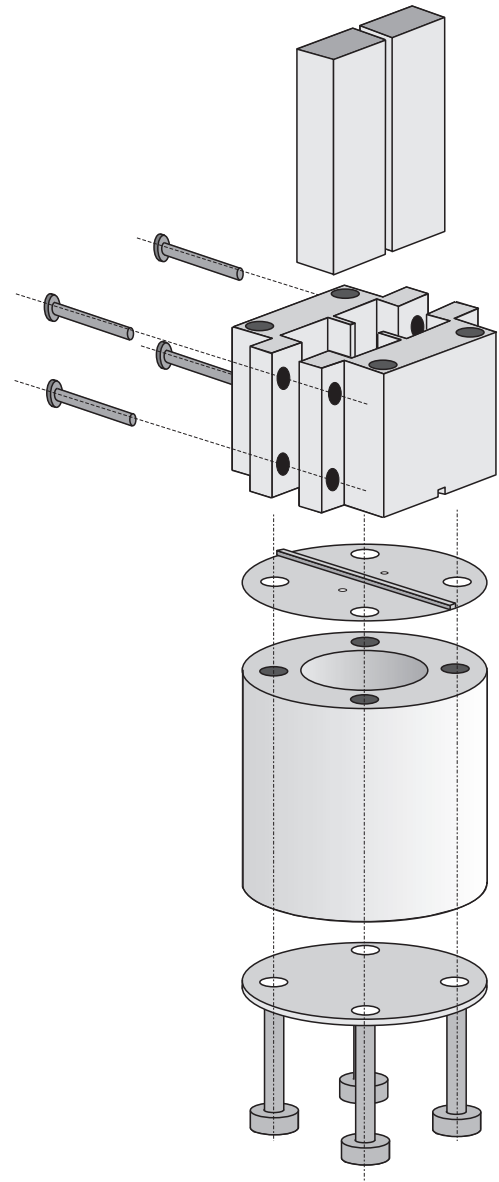,width=65mm}} \bigskip
\caption{Schematic of the axial cavity construction. See text for
detailed description.} \label{Fig. 7}
\end{figure}

\clearpage

\begin{figure}
\centerline{\epsfig{figure=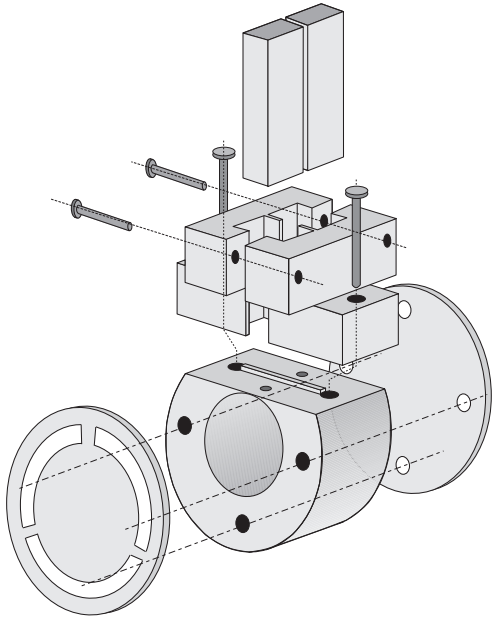,width=65mm}} \bigskip
\caption{Schematic of the transverse cavity construction. See text
for detailed description.} \label{Fig. 8}
\end{figure}

\break

\begin{figure}
\centerline{\epsfig{figure=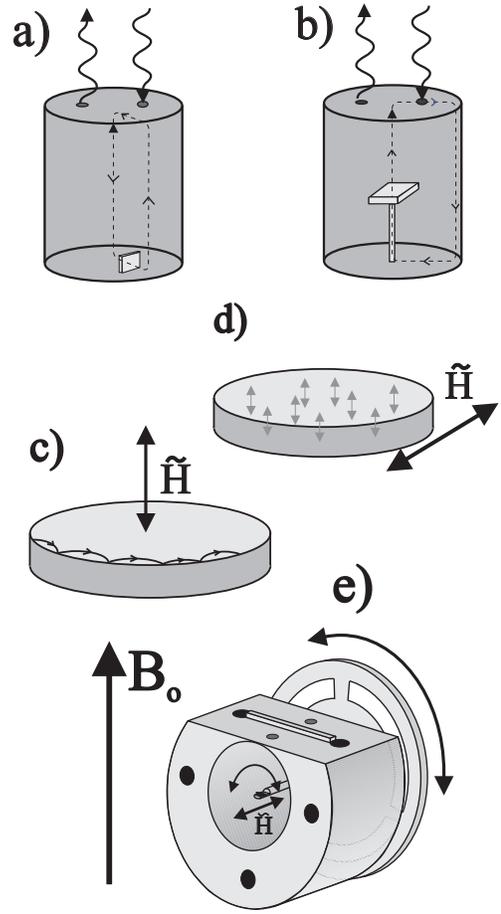,width=65mm}} \bigskip
\caption{a) The end plate cavity perturbation configuration,
showing the position of the sample within a TE011 cavity and the
radial $\tilde{H}$-field at the position of the sample. b) The
quartz pillar configuration, showing the sample at the heart of a
TE011 cavity atop a quartz pillar, and the axial $\tilde{H}$-field
at the position of the sample. c) Excitation of in-plane currents
at the edges of a Q2D conductor. d) Excitation of inter-layer
currents within the bulk of a Q2D conductor. In both c) and d),
the low conductivity direction is parallel to the normal to the
disc shaped sample. e) Sample rotation for the quartz pillar
configuration in the transverse cavity $-$ note that the
orientation of the axial AC $\tilde{H}$-field does not change
relative to the sample upon rotation.} \label{Fig. 9}
\end{figure}

\clearpage

\begin{figure}
\centerline{\epsfig{figure=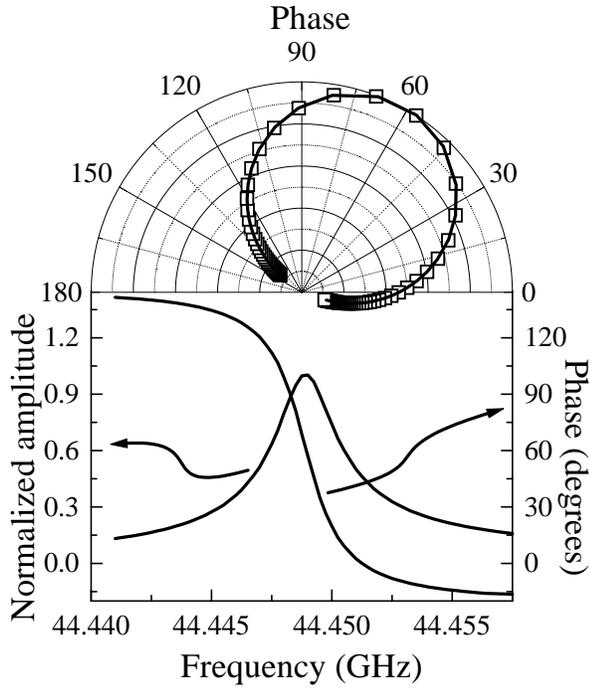,width=80mm}} \bigskip
\caption{A real TE011 resonance (raw data at 4.2 K) for a loaded
axial cavity (cavity A). The upper panel shows the resonance
circle in the complex plane - the squares are raw data and the
solid line is a fit. The lower panel shows the linear amplitude
and phase variation as a function of frequency.} \label{Fig. 10}
\end{figure}


\begin{figure}
\centerline{\epsfig{figure=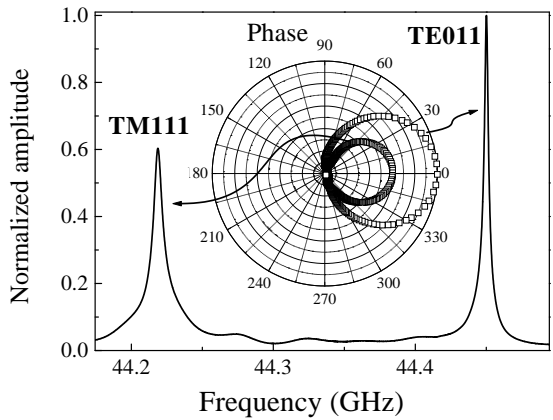,width=80mm}} \bigskip
\caption{Lifting the degeneracy of the TE011 and TM111 modes (T =
4.2 K) in cavity A. The main part of the figure shows the two
resonances (linear scale) separated by 230 MHz. The inset shows
the circles in the complex plane corresponding to the two
resonances.} \label{Fig. 11}
\end{figure}

\begin{figure}
\centerline{\epsfig{figure=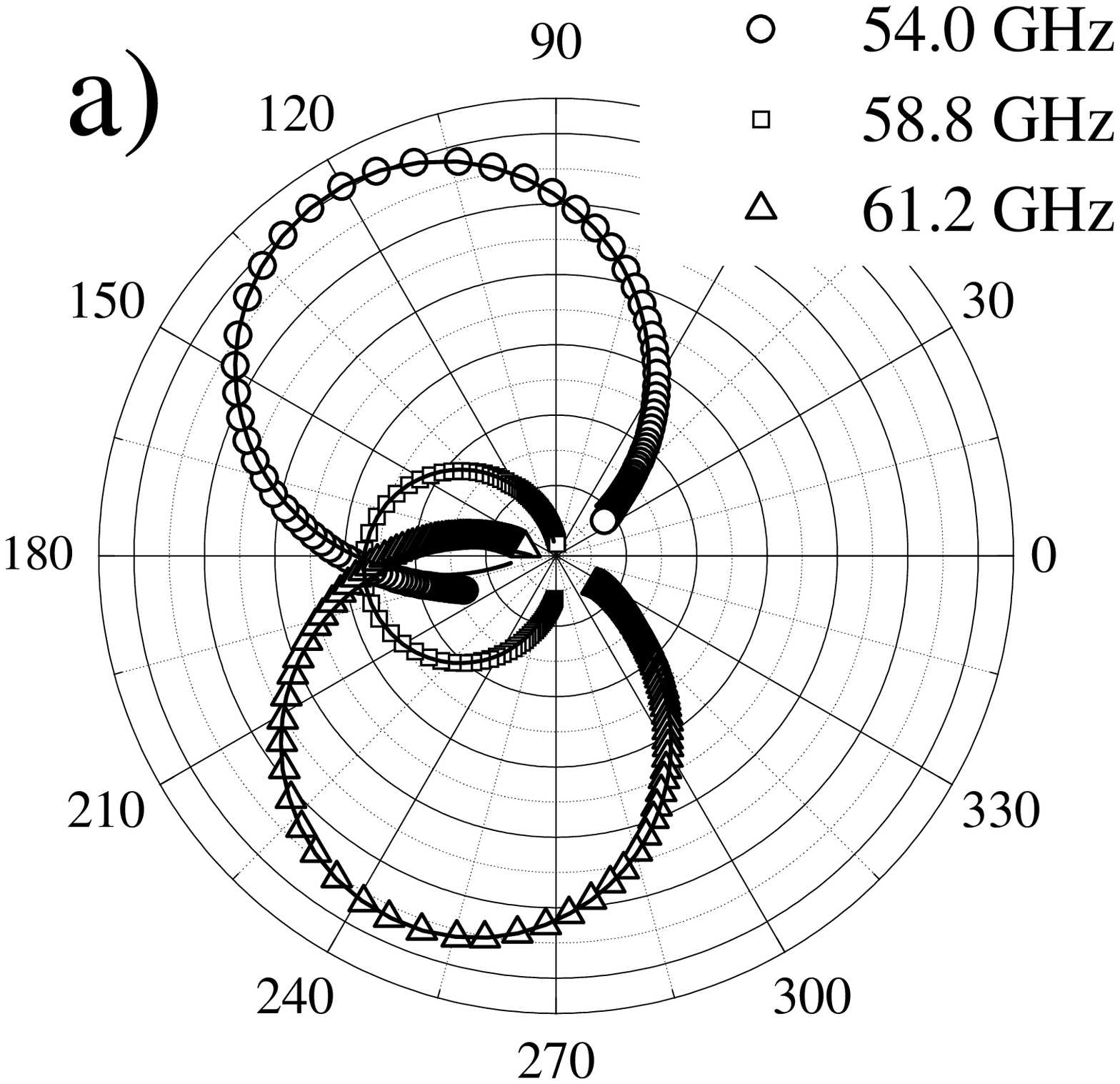,width=75mm}}
\centerline{\epsfig{figure=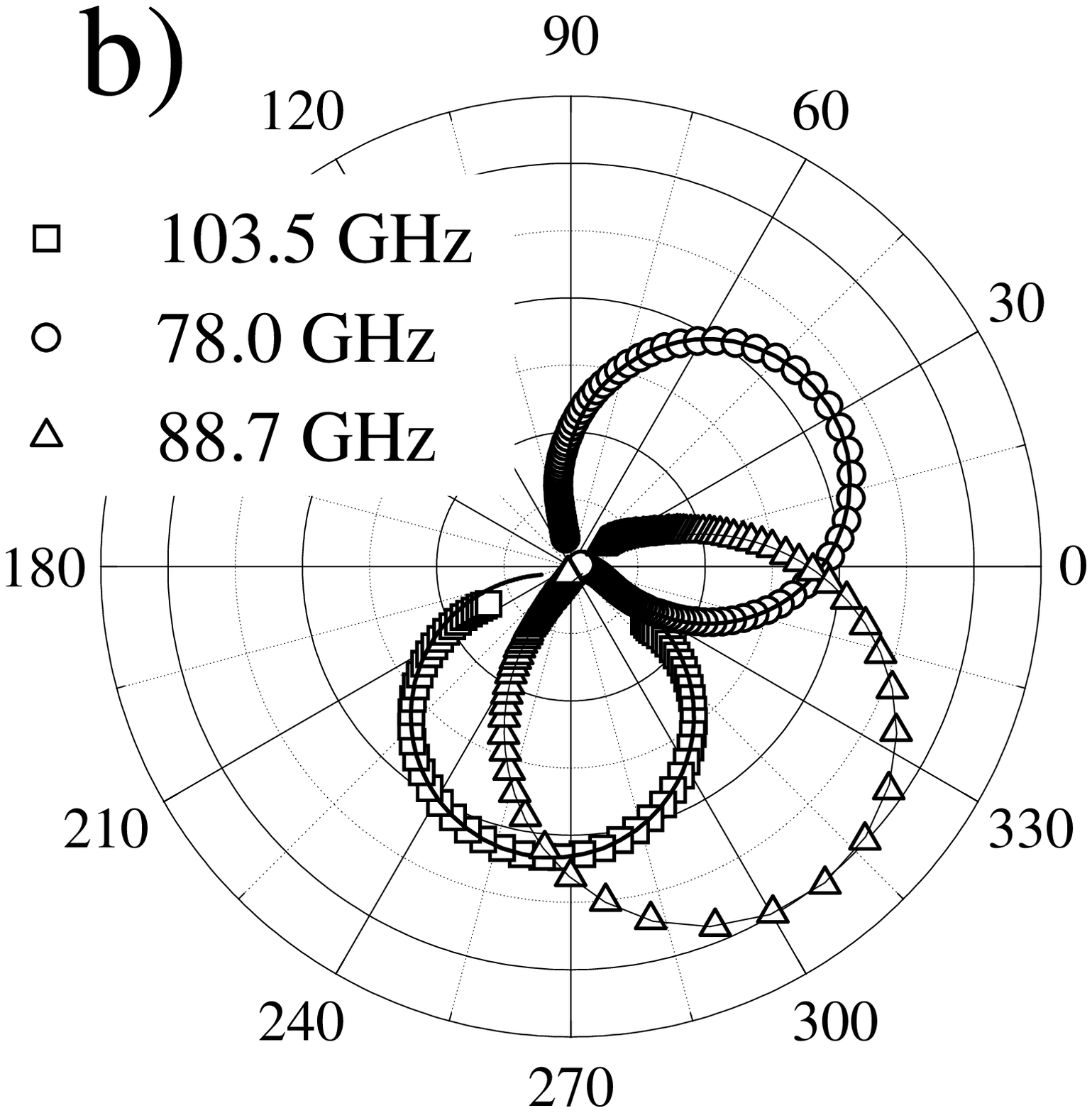,width=75mm}}
\centerline{\epsfig{figure=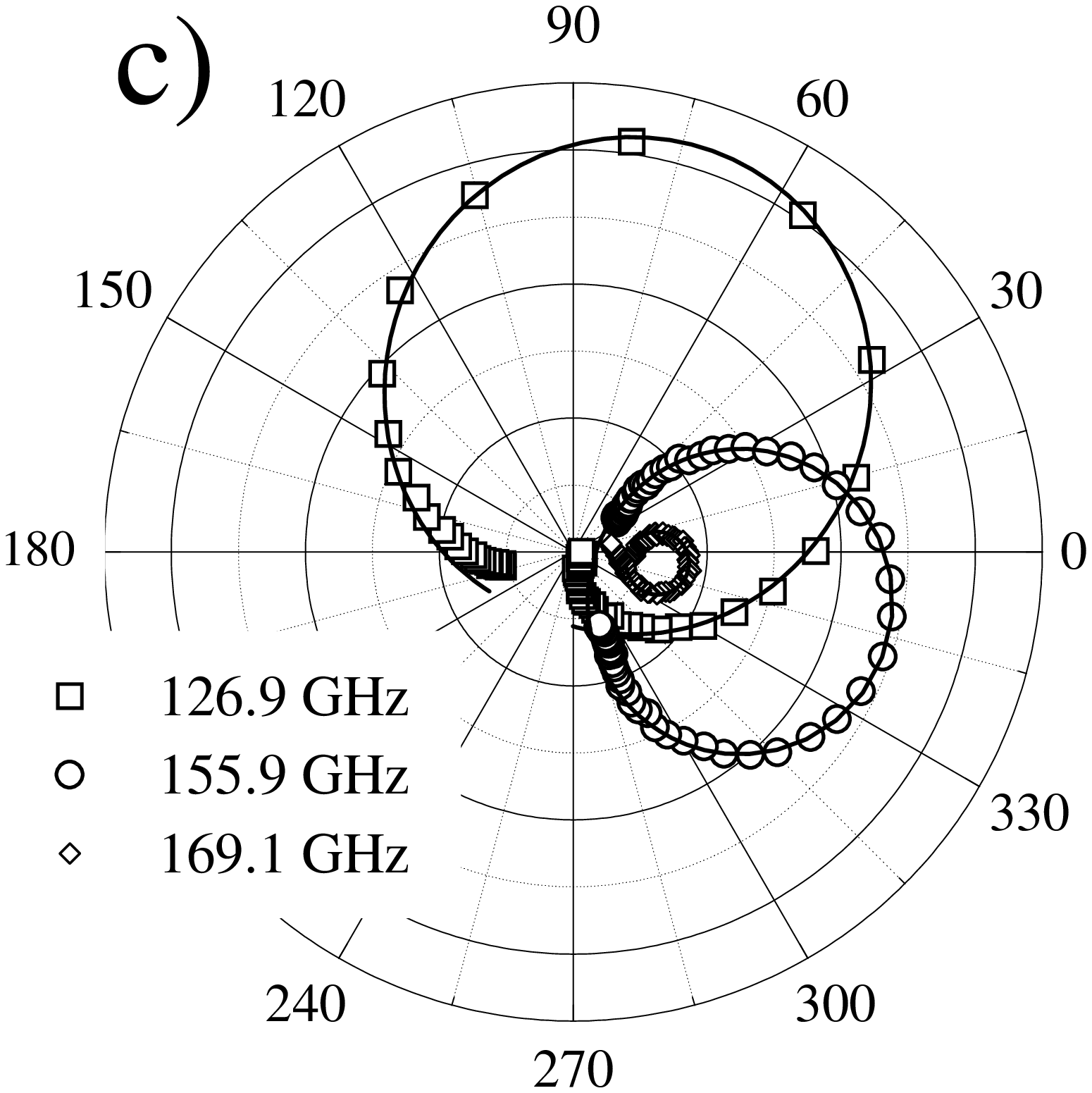,width=75mm}} \bigskip
\caption{Various cavity resonances seen as circles in the complex
plane, obtained in a) V-band, b) W-band and c) D-band. The points
are raw 4.2 K data and the solid lines represent Lorentzian fits
to the data. See table 2 for more details about each of the
resonances.} \label{Fig. 12}
\end{figure}

\clearpage

\begin{figure}
\centerline{\epsfig{figure=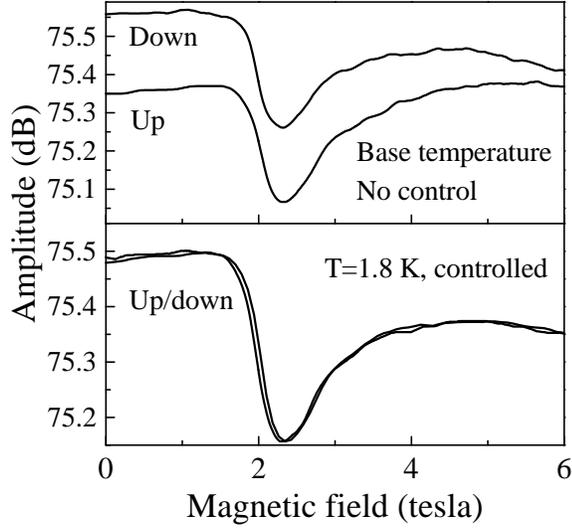,width=80mm}} \bigskip
\caption{A comparison between magnetic field sweeps with the
temperature unlocked (upper panel) and actively controlled (lower
panel). The broad dip corresponds to the Fe$^{3+}$ resonance seen
in Fig. 6. Note: this data was obtained prior to Au plating the
lower SS section of the probe.} \label{Fig. 13}
\end{figure}

\begin{figure}
\centerline{\epsfig{figure=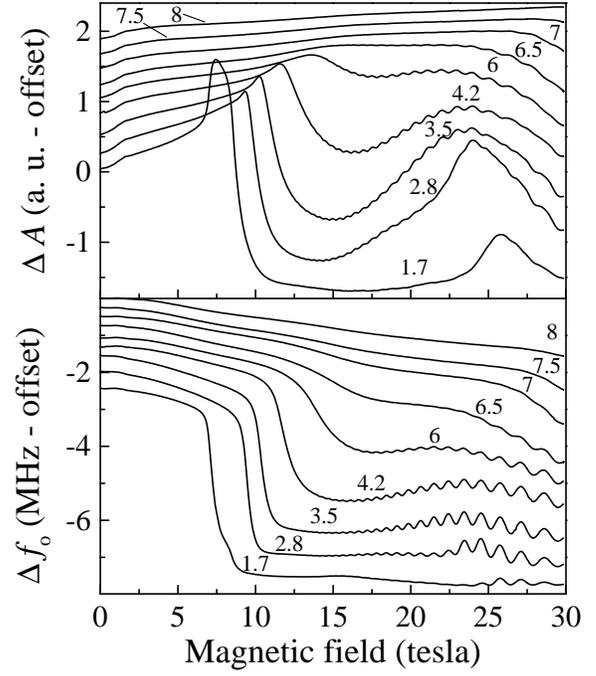,width=80mm}} \bigskip
\caption{Dissipative ($\Delta A$) and reactive ($\Delta f_o$)
responses of (TMTSF)$_2$ClO$_4$ in a magnetic field. The data were
obtained in an axial TE011 cavity excited at 47.2 GHz. The
temperatures are indicated in kelvin in the figure, and the traces
have been offset for the sake of clarity.} \label{Fig. 14}
\end{figure}

\clearpage

\begin{figure}
\centerline{\epsfig{figure=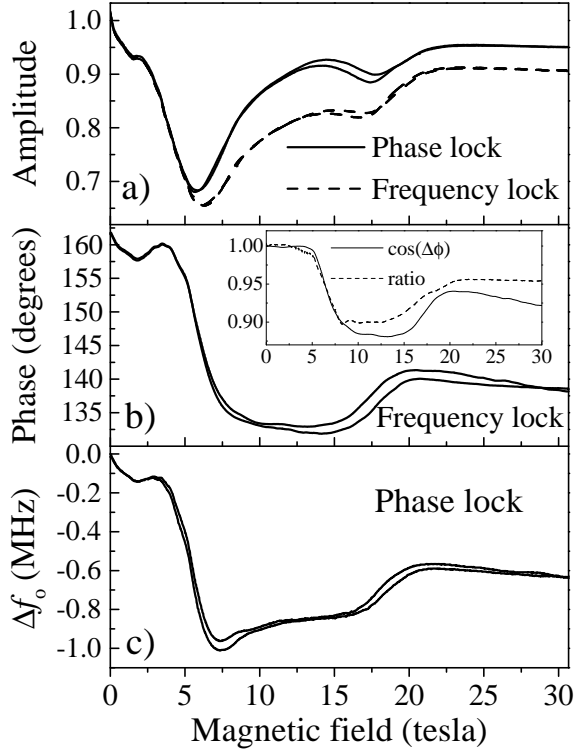,width=80mm}} \bigskip
\caption{a) A comparison between the amplitude variation versus
magnetic field obtained using frequency and phase locking
techniques. The sample is the purple bronze $\eta
-$Mo$_4$O$_{11}$, and the observed changes are due to the
excitation of inter-layer currents in this quasi-two-dimensional
conductor. b) Shows the phase variation during the frequency
locked measurement (absolute microwave frequency fixed at 44.272
GHz), while c) shows the frequency variation during the phase
locked measurement (absolute microwave phase across the sample
probe locked). The inset to b) compares the ratio of the
amplitudes obtained for the two techniques in a), with the cosine
of the phase change in the main part of b). The good
correspondence between these quantities confirms that the
difference between the traces in a) can be attributed to the large
phase shift during the frequency locked measurement. Thus, a phase
lock should be applied in this situation.} \label{Fig. 15}
\end{figure}

\begin{figure}
\centerline{\epsfig{figure=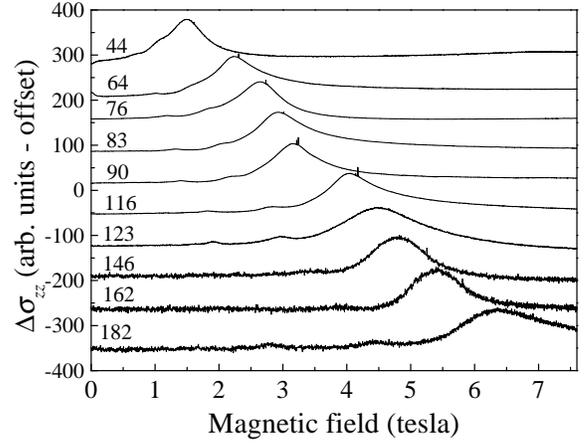,width=80mm}} \bigskip
\caption{Frequency dependence of the Periodic Orbit Resonances
observed through the inter-layer conductivity ($\sigma _{zz}$) in
$\alpha -$(BEDT-TTF)$_2$TlHg(NCS)$_4$, when mounted in a TE011
axial cavity in the end plate configuration; the temperature is
1.8 K. The data appear noisier at higher frequencies due to the
reduced dynamic range of the spectrometer and due to the reduced
sensitivity (lower $Q-$values) of the higher order cavity modes.
The resonances have been scaled so as to appear the same strength,
and have been offset for the sake of clarity. The sharp spikes
observed slightly above the main peak positions are Electron
Paramagnetic Resonance signals due to a DPPH sample placed in the
cavity.} \label{Fig. 16}
\end{figure}


\begin{figure}
\centerline{\epsfig{figure=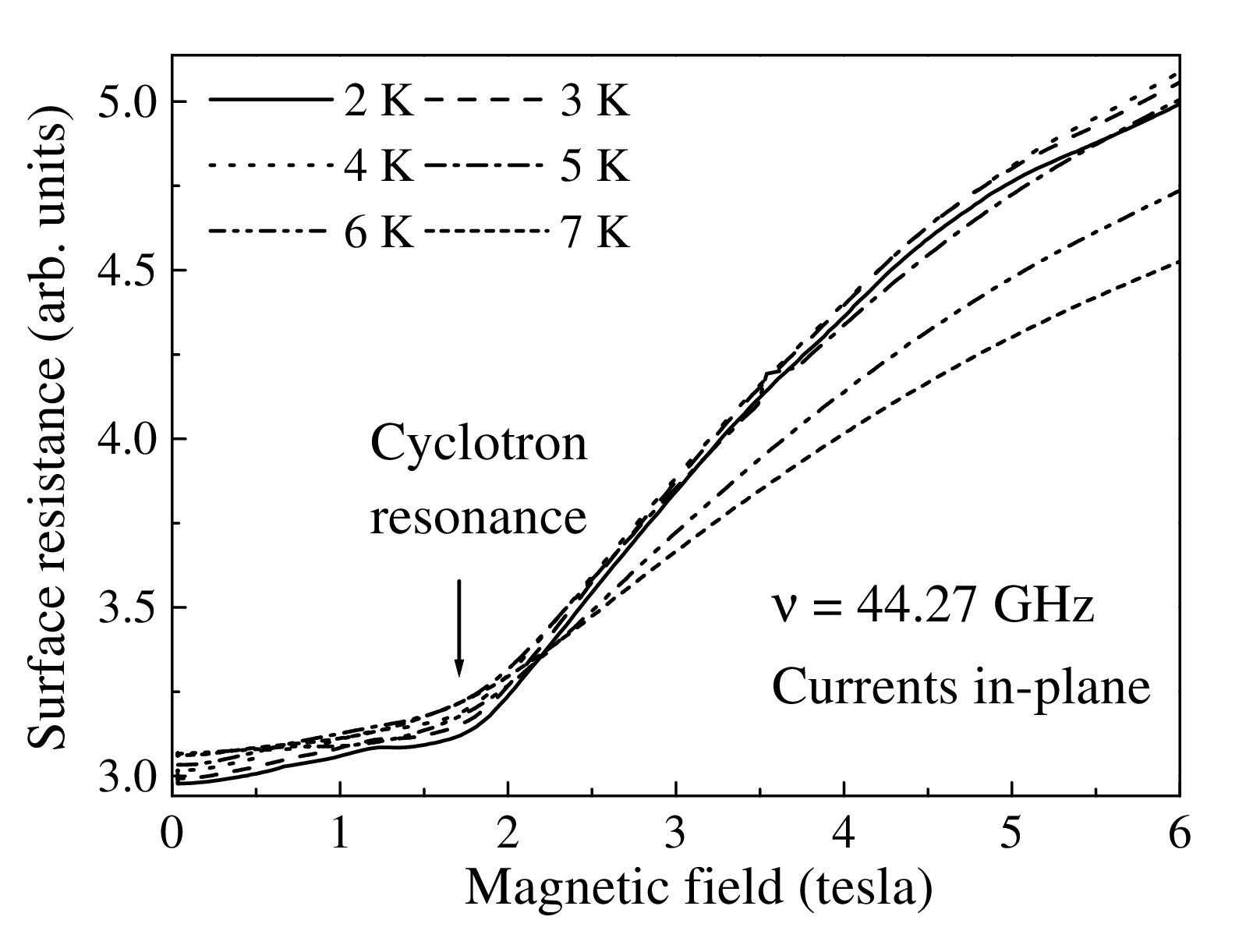,width=80mm}} \bigskip
\caption{Conventional cyclotron resonances seen through the
in-plane surface resistance for the same sample as in Fig. 16,
this time mounted in a TE011 axial cavity in the quartz pillar
configuration; the frequency is 44.27 GHz and the temperatures are
indicated in the figure. The resonances are observed as sharp
inflections which weaken as the temperature increases.}\label{Fig.
17}
\end{figure}

\end{twocolumn}


\begin{references}
\bibitem{to1thz}This may be extended to 1 THz through the association
with Gunn oscillators.
\bibitem{mvnatypes}The numbers after MVNA 8-350 refer to the precise configuration
of the MVNA. In this case it is the configuration with a single
source and a single detector, and with a dual channel receiver,
allowing vector detection of two microwave frequencies at the same
time, or the vector detection of a signal and its derivative
versus magnetic field.
\bibitem{bore}The warm bore diameter of these magnets is 32mm. For
measurements at pumped liquid helium temperatures, space is
restricted to a 19mm diameter bore.
\bibitem{kappa}The thermal conductivities at 77 K for SS,
electrolytic Cu and Ag are 0.08, 5.5 and 3.1 W cm$^{-1}$K$^{-1}$,
respectively.
\bibitem{wall}SS waveguide is manufactured with a 0.51mm wall, as
opposed to 1.02mm for Cu and Ag.
\bibitem{rho}This is due to the high electrical resistivity of SS compared
to Ag and Cu. The electrical resistivities at 300 K for SS,
electrolytic Cu and Ag are 72, 1.67 and 1.59 m$\Omega$ cm,
respectively.
\bibitem{cuag}The reason we use both Cu and Ag in the probe simply has
to do with availability at the time of construction. Thus, at the lower end
of the probe, one of the guides is Cu and the other is Ag.
\bibitem{phirot}More precisely, 1m of cable compensates for 1.18m in air
(vacuum). However, in waveguides close to cut-off, this conversion factor may
not be reliable.
\bibitem{ease}Although our goal is to be able to measure at many different
frequencies with a single cavity, different cavities are needed
for different experiments, $e.g$. a long cavity will provide many
closely spaced modes within a narrow frequency range, whereas a
short cavity would spread these modes out over a broad frequency
range.
\bibitem{hybrid}This will shortly increase to 45 T when the hybrid magnet comes on line.
\bibitem{vr}Although the VR detects at 10 kHz, it is common to average
this signal over about 50 cycles, resulting in a detection frequency of
about 200 Hz.
\end{references}
\end{document}